\documentclass[preprint,authoryear]{elsarticle}
\usepackage[utf8]{inputenc}
\usepackage{amsmath}
\usepackage{mathtools}
\DeclareGraphicsExtensions{.pdf}
\usepackage{lineno}
\usepackage{hyperref}
\usepackage{multirow}
\usepackage{multicol}
\usepackage[table]{xcolor}
\usepackage{booktabs}
\usepackage{tikz}
\usetikzlibrary{calc, positioning, backgrounds, arrows}
\tikzset{every picture/.style={/utils/exec={\sffamily}}}

\newcommand{\lite}{\textsubscript{Lite}}
\newcommand{\iclbeta}{\textsubscript{Beta}}
\newcommand{\ac}{\textsubscript{AC}}
\newcommand{\icmsf}{IC\_MARC\textsubscript{SF}}
\newcommand{\icmef}{IC\_MARC\textsubscript{EF}}
\newcommand{\rmtp}{\text{TP}}
\newcommand{\rmfp}{\text{FP}}
\newcommand{\rmtn}{\text{TN}}
\newcommand{\rmfn}{\text{FN}}
\DeclareMathOperator*{\argmax}{arg\,max}

\begin{document}
	
	\begin{frontmatter}
		
		\title{ICLabel: An automated electroencephalographic independent component classifier, dataset, and website}
		
		\author[sccn,ece]{Luca Pion-Tonachini\corref{mycorrespondingauthor}}
		\cortext[mycorrespondingauthor]{Corresponding author}
		\ead{lpionton@ucsd.edu}
		\author[ece,prl]{Ken Kreutz-Delgado}
		\ead{kreutz@eng.ucsd.edu}
		\author[sccn]{Scott Makeig}
		\ead{smakeig@ucsd.edu}
		
		\fntext[fn1]{Declaration of interest: none.}

		\address[sccn]{Swartz Center for Computational Neuroscience, University of California San Diego, 9500 Gilman Drive, La Jolla CA 92093, USA}
		\address[ece]{Department of Electrical and Computer Engineering, University of California San Diego, 9500 Gilman Drive, La Jolla CA 92093, USA}
		\address[prl]{Pattern Recognition Laboratory, University of California San Diego, 9500 Gilman Drive, La Jolla CA 92093, USA}
		
		\begin{abstract}
			The electroencephalogram (EEG) provides a non-invasive, minimally restrictive, and relatively low-cost measure of mesoscale brain dynamics with high temporal resolution. 
			Although signals recorded in parallel by multiple, near-adjacent EEG scalp electrode channels are highly-correlated and combine signals from many different sources, biological and non-biological, independent component analysis (ICA) has been shown to isolate the various source generator processes underlying those recordings. 
			Independent components (IC) found by ICA decomposition can be manually inspected, selected, and interpreted, but doing so requires both time and practice as ICs have no order or intrinsic interpretations and therefore require further study of their properties.
			Alternatively, sufficiently-accurate automated IC classifiers can be used to classify ICs into broad source categories, speeding the analysis of EEG studies with many subjects and enabling the use of ICA decomposition in near-real-time applications.
			While many such classifiers have been proposed recently, this work presents the ICLabel project comprised of (1) the \emph{ICLabel dataset} containing spatiotemporal measures for over 200,000 ICs from more than 6,000 EEG recordings and matching component labels for over 6,000 of those ICs, all using common average reference,
			(2) the \emph{ICLabel website} for collecting crowdsourced IC labels and educating EEG researchers and practitioners about IC interpretation, and (3) the automated \emph{ICLabel classifier}, freely available for MATLAB. 
			The ICLabel classifier improves upon existing methods in two ways: by improving the accuracy of the computed label estimates and by enhancing its computational efficiency. 
			The classifier outperforms or performs comparably to the previous best publicly available automated IC component classification method for all measured IC categories while computing those labels ten times faster than that classifier as shown by a systematic comparison against other publicly available EEG IC classifiers.
		\end{abstract}
		
	
		
	\end{frontmatter}
	
	\section{Introduction and Overview}
	
	Electroencephalography (EEG) is a non-invasive, functional brain-activity recording modality with high temporal resolution and relatively low cost. 
	Despite these benefits, an unavoidable and potentially confounding issue is that EEG recordings mix activities of more sources than just the participant’s brain activity. Each EEG electrode channel collects a linear mixture of all suitably projecting electrical signals, some of them not originating from the cortex or even from other biological sources. The relative proportions of those mixtures depend on the positions and orientations of the signal generators and the electric fields they produce relative to each recording channel, which always records the difference between activity at two or more scalp electrodes.
	This mixing process applies to brain activity as well. Far-field electrical potentials from regions of locally-coherent cortical field activity will not only reach the closest EEG electrodes, but nearly the whole electrode montage to varying degrees \citep{delorme2012independent, brazier1966study}.
	Independent component analysis (ICA) \citep{doi:10.1162/neco.1995.7.6.1129,lee1999independent,palmer2008newton} has been shown to unmix and segregate recorded EEG activity into maximally independent generated signals \citep{makeig1996independent,jung1998extended,delorme2012independent}.
	By assuming that the original, unmixed source signals are spatially stationary and statistically independent of each other, and that the mixing occurs linearly and instantaneously, ICA simultaneously estimates both a set of linear spatial filters that unmix the recorded signals and the source signals that are the products of that linear unmixing.
	
	A typical multichannel EEG recording contains electrical far-field signals emanating from different regions of the participant’s brain in which cortical tissue generates synchronous electrical potentials \citep{malmivuo1995bioelectromagnetism}. Further potentials arise in the subject’s eyes that project differently to the scalp as their eyes rotate. Electromyographic (EMG) activity associated with any muscle contractions strong and near enough to the electrodes are also summed into the recorded EEG signals. Even electrocardiographic (ECG) signals originating from the participant's heart can appear in scalp EEG recordings. Entirely non-biological signals such as 50-Hz or 60-Hz oscillations induced by alternating current electrical fixtures such as fluorescent lights may also contribute to the recorded EEG. The electrodes themselves can introduce artifacts into the recorded signals when the electrode-skin interface impedance is large or unstable. All of these electrical fields and signal artifacts are combined to form the instantaneous, linear mixture of signals recorded in each electrode channel. However, the source signals themselves are largely generated independently and should not have any consistent instantaneous effect upon one another, justifying the use of ICA decomposition.
	
	Though useful, the application of ICA to EEG data introduces two problems: (1) sensitivity to noise and artifacts and (2) ambiguity of the ICA results.
	If too many artifacts are present in an EEG recording, or even just a few with extreme amplitudes, the ICA solution found may be unusable or noisy, comprised of crudely defined independent components (IC), each summing poorly unmixed source signals. This problem can be mitigated through adequate signal preprocessing prior to applying ICA and, as many effective preprocessing pipelines already exist \citep{10.3389/fninf.2015.00016, mullen2013real}, this work does not address preprocessing. Instead, we address the issue of resolving ambiguity in ICA solutions, a problem which results from the fact that ICA is an unsupervised learning method. As ICA does not consider any signal or event annotations in conjunction with the EEG data, any structure present in the ICA solution thereby lacks explicit labels. Consequently, the raw ICA output is an unordered and unlabeled set of ICs. One common step towards organizing the results is to standardize the IC scalp projection norms and order ICs by descending time-series activity power. Even so, the provenance of each IC signal is difficult to determine without sufficient training and time dedicated to manual inspection.
	An automated solution to determining IC signal categories, referred to as IC classification or IC labeling, would aid the study and use of EEG data in four ways:
	\begin{enumerate}
		\item Provide consistency in the categorization of ICs. 
		\item Expedite IC selection in large-scale studies.
		\item Automate IC selection for real-time applications including brain-computer interfaces (BCI).
		\item Guide IC selection for people lacking the necessary training and help them to learn through examples.
	\end{enumerate}
	
	
	
	This work presents a new IC classifier, along with the dataset used to train and validate that classifier and the website used to collect crowdsourced IC labels for the dataset. The classifier is referred to as the ICLabel classifier while the dataset and website are referred to as the ICLabel dataset and ICLabel website, respectively. The process for creating and validating the ICLabel classifier began with the creation of the ICLabel dataset and website, as the website was used to annotate the dataset needed to make the classifier.
	
	The first step was to create the ICLabel training set by collecting examples of EEG ICs and pairing them with classifications of those ICs.
	The ICLabel website (\url{https://iclabel.ucsd.edu/tutorial})was designed with the express purpose of generating these IC labels for ICs that had no prior annotations. The website also functions as an educational tool as well as a crowdsourcing platform for accumulating redundant IC labels from website users. These redundant labels are then combined, using a crowd  labeling (CL) algorithm, to generate probabilistic labels for the training set. In addition to the ICLabel training set, we also constructed a second ICLabel expert-labeled test set containing additional ICs not present in the training set, used for classifier validation. 
	
	With this foundation in place, the next step was to create and validate the ICLabel classifier. To do so, multiple candidate classifiers were trained using the ICLabel training set and the final ICLabel classifier was modeled after the candidate classifier that best performed on the cross-validated training set. Once trained on the ICLabel training set, the ICLabel classifier was validated against other publicly available IC classifiers on the ICLabel expert-labeled test set. The final products of this process are the ICLabel classifier, dataset, and website, all of which are freely available online. The classifier may be downloaded through the EEGLAB extensions manager under the name ICLabel or may be downloaded directly from \url{https://github.com/sccn/ICLabel}. The ICLabel dataset may be downloaded from \url{https://github.com/lucapton/ICLabel-Dataset} and the educational ICLabel website is accessible at \url{https://iclabel.ucsd.edu/tutorial}.
	
	\section{Background}
	
	\subsection{EEG Component Interpretation} 
	\label{sec:ictypes}
	When a signal generator produces electric fields with a stable spatial projection pattern across the recording electrodes, ICA decomposition may capture that activity in one IC.
	Perfect separation of source signals is not always possible and, often, is difficult to verify without concurrent invasive recordings. Suboptimal signal unmixing can happen because of poor ICA convergence due to an insufficient amount of clean data or excessive artifacts and noise in the data. Some source signals cannot be fully described in one IC, as when signal source projections are not spatially stationary. However, due to the iterative nature of the convergence of ICA algorithms, most ICs primarily account for one specific source signal, even when some sources are not perfectly separated \citep{hsu2014online}. To simplify further discussion, rather than referring to, for example, ``primarily brain-related" or ``non-brain-related" ICs, ICs accounting predominantly for activity originating within the brain will be referred to as ``Brain ICs". This verbal denotation can be generalized to any number of IC categories, the definitions of which are provided in Section \ref{sec:ictypes}. While this denotation is simpler to read and write, it also hides the possibility of complexities and imperfections in the ICs and in the signals they describe. It is therefore important that the reader not forget the possible intricacies masked by this simple nomenclature.
	
	\subsection{Prior Methods}
	\label{sec:prior_methods}
	Several other attempts to automatically solve the IC classification problem have been made publicly available. A recent and largely comprehensive summary of those methods can be found in the introduction of \citet{tamburro2018new}. For our purposes, we only consider and compare methods and their supporting algorithms that are (1) publicly available, (2) do not require any information beyond the ICA-decomposed EEG recordings and generally available meta-data such as electrode locations, and (3) have at minimum a category for Brain ICs as defined in Section \ref{sec:ictypes}.
	This excludes IC classification methods that have not released the trained classifiers, classifiers that only classify certain non-brain artifacts, and methods that require additional recordings such data from an electrooculogram (EOG), ECG, electromyogram (EMG), or accelerometer.
	
	Provided the first two constraints hold, a direct comparison of all accessible methods on a common collection of datasets becomes possible and is presented in Section \ref{sec:result-test}.
	EEG IC classifiers that matched the above criteria are summarized here:
	\begin{itemize}
		\item \textbf{MARA} \citep{winkler2011automatic,winkler2014robust} is an IC classifier that estimates the probability of ICs being either (non-brain) artifactual or Brain ICs. It uses a regularized LDA model trained on 23 EEG recordings consisting of 690 ICs. All ICs were labeled by two experts. All recordings used the same experimental paradigm.
		\item \textbf{ADJUST} \citep{mognon2011adjust} classifies ICs into five discrete categories, three of which are related to eye activity. Its feature-specific thresholds were learned from 20 EEG recordings for a single experimental paradigm.
		\item \textbf{FASTER} \citep{nolan2010faster} was intended as a full processing pipeline that cleans unprocessed, raw EEG data. Only the portion that classifies ICs is considered here. FASTER labels an IC as ``artifactual" if any of the features it calculates deviates from the dataset average by more than three standard deviations.
		\item \textbf{SASICA} \citep{chaumon2015practical} performs semi-automatic classification based on features from MARA, FASTER, and ADJUST plus additional features. SASICA was primarily intended as an educational tool to help users learn how to manually label ICs. It uses feature-specific thresholds to determine which ICs should be rejected, presumably keeping only Brain ICs for further analysis. When operating automatically, SASICA uses thresholds between two to four standard deviations from the dataset average. Alternatively, thresholds may be manually chosen.
		\item \textbf{IC\_MARC} \citep{doi:10.1111/psyp.12290} uses a multinomial logistic regression model trained on 46 EEG recordings comprising 8023 ICs and two experimental paradigms. The associated publication describes two versions. In the first, the features were selected using two-level cross-validation over a larger initial set of features, referred to as the established feature set (\icmef{}). The second version uses selected spatial features and, while originally intended for short recordings, appears to work better in practice, and is referred to below as the spatial feature set (\icmsf{}). Both versions compute probabilistic labels over six classes, two of which are related to eye activity.
	\end{itemize}
	
	Despite the existence of these IC classification methods and others, there remains room for improvement by increasing output \emph{descriptiveness}, \emph{accuracy}, and \emph{efficiency}, terms which are defined as follows.
	An IC classifier can be said to be more \emph{descriptive} if it can differentiate between a larger number of useful IC categories and if the classifications provided are probabilistic across all relevant categories rather than discrete, single-category determinations. In the case of an ambiguous EEG component with hard labels, there is no recourse to convey that ambiguity. If a discrete classifier produces an incorrect component label, there is also no way to find the next best category from the discrete classification. FASTER, ADJUST, and SASICA are examples of classifiers that produce discrete classifications. This is discussed further in Section \ref{sec:compositional_labels}.
	
	\emph{Accuracy} refers not only to classifier performance on the same type of data it was trained on, but how well that classifier's performance generalizes across all EEG data, independent of experiment, recording environment, amplifier, electrode montage, preprocessing pipeline, etc. Though measuring performance across all possible datasets is infeasible, computing performance across multiple experiments and recording conditions should be a minimum requirement. The previous methods listed above used one or two experiment types with the exception of SASICA and MARA which used more. Furthermore, because even expert human IC classifiers often disagree \citep{chaumon2015practical,doi:10.1111/psyp.12290} it is important to find a consensus among multiple labelers. This is a matter that many of the prior projects handled well, although some did not explicitly report how many labelers, expert or otherwise, were used.
	
	\emph{Efficiency} refers to the computational load and speed of extracting the required IC features and computing IC classifications. While generally beneficial, efficiency is only situationally important. Specifically, efficiency is paramount when IC classification is desired for online streaming data. Without a computationally efficient classifier, the delay incurred when classifying ICs may negate any utility gained through obtaining the classifications. In offline cases, efficiency is merely a matter of convenience and, possibly, of cost.
	
	\subsection{The ICLabel Project}
	The ICLabel project provides improved classifications based on the aforementioned desirable qualities of an EEG IC classifier. To be sufficiently \emph{descriptive}, the ICLabel classifier computes IC class probabilities across seven classes as described below. To achieve \emph{accuracy} across EEG recording conditions, the ICLabel dataset used to train and evaluate the ICLabel classifier encompasses a wide variety of EEG datasets from a multitude of paradigms. These example ICs are paired with component labels collected through the ICLabel website from hundreds of contributors. Finally, to maintain sufficient computational \emph{efficiency}, relatively simple IC features are used as input to an artificial neural network architecture (ANN) that, while slow to train, computes IC labels quickly. The end result is made freely and easily available through the ICLabel plug-in for the EEGLAB software environment \citep{delorme2004eeglab,delorme2011eeglab}.
	
	The seven IC categories addressed in this work are:
	\begin{itemize}
		\item \textbf{Brain} ICs contain activity believed to originate from locally synchronously activity in one (or sometimes two well-connected) cortical patches. The cortical patches are typically small and produce smoothly varying dipolar projections onto the scalp. Brain ICs tend to have power spectral densities with inversely related frequency and power and, often, exhibit increased power in frequency bands between 5 and 30 Hz. See Figure \ref{fig:website} for an example of a Brain IC.
		\item \textbf{Muscle} ICs contain activity originating from groups of muscle motor units (MU) and contain strong high-frequency broadband activity aggregating many MU action potentials (MUAP) during muscle contractions and periods of static tension. 
		These ICs are effectively surface EMG measures recorded using EEG electrodes. They are easily recognized by high broadband power at frequencies above 20-30 Hz. Often times they can appear dipolar like Brain ICs, but as their sources are located outside the skull, their dipolar pattern is much more localized than for Brain sources.
		\item \textbf{Eye} ICs describe activity originating from the eyes, induced by the high metabolic rate in the retina that produces an electrical dipole (positive pole at the cornea, negative at the retina) \citep{malmivuo1995bioelectromagnetism}. 
		Rotating the eyes shifts the projection of this standing dipole to the frontal scalp.
		Eye ICs can be further subdivided into ICs accounting for activity associated with horizontal eye movements and ICs accounting for blinks and vertical eye movements. Both have scalp projections centered on the eyes and show clear quick or sustained ``square" DC-shifts depending on whether the IC is describing blinks or eye movements respectively.
		\item \textbf{Heart} ICs, though more rare, can be found in EEG recordings. They are effectively electrocardiographic (ECG) signals recorded using scalp EEG electrodes. They are recognizable by the clear QRS-complexes \citep{malmivuo1995bioelectromagnetism} in their time series and often have scalp projections that closely approximate a diagonal linear gradient from left-posterior to right-anterior. Heart ICs can rarely have localized scalp projections if an electrode is placed directly above a superficial vein or artery.
		\item \textbf{Line Noise} ICs capture the effects of line current noise emanating from nearby electrical fixtures or poorly grounded EEG amplifiers. They are immediately recognizable by their high concentration of power at either 50 Hz or 60 Hz depending on the local standard. These effects can only be well separated if the line noise interference is spatially stationary across the EEG electrodes. Otherwise, it is unlikely that a single IC will be able to describe the line noise activity. Instead, several or even all components may be contaminated to varying degrees. 
		\item \textbf{Channel Noise} ICs indicate that some portion of the signal recorded at an electrode channel is already nearly statistically independent of those from other channels. These components can be produced by high impedance at the scalp-electrode junction or physical electrode movement, and are typically an indication of poor signal quality or large artifacts affecting single channels. If an ICA decomposition is primarily comprised of this IC category, that is a strong indication that the data has received insufficient preprocessing. In this paper, ``Channel Noise" will sometime be shortened to ``Chan Noise".
		\item \textbf{Other} ICs, rather than being an explicit category, act as a catch-all for ICs that fit none of the previous types. These primarily fall into two categories: ICs containing indeterminate noise or ICs containing multiple signals that ICA decomposition could not separate well. For ICA-decomposed high-density EEG recordings (64 channels and above), the majority of ICs typically fall into this category.
	\end{itemize}


	\section{Materials and Methods}
	
	\subsection{ICLabel Dataset and Website}
	\label{sec:data}
	The ICLabel training set used to train the ICLabel classifier currently has been drawn from 6,352 EEG recordings collected from storage drives at the Swartz Center for Computational Neuroscience (SCCN) at UC San Diego (\url{https://sccn.ucsd.edu}). These datasets encompass a portion of the experiments recorded at the SCCN or brought to the SCCN by visiting researchers since 2001. In aggregate, these recordings include a total of 203,307 unique ICs; none of which had standardized IC classification metadata and were therefore effectively unlabeled for the purposes of this project. Prior to computing features, each dataset was converted to a common average reference \citep{dien1998issues}. For each IC, the ICLabel training set includes a set of standard measures: a scalp topography, median power spectral density (PSD) and autocorrelation function, and single and bilaterally symmetric equivalent current dipole (ECD) model fits, plus features used in previously published classifiers (ADJUST, FASTER, SASICA, described in Section \ref{sec:prior_methods}). These features potentially provide an IC classifier with information contributory to computing accurate component labels.
	
	\subsubsection{IC Features Descriptions}
	
	Scalp topographies are a visual representation of how IC activity projects to the subject’s scalp by interpolating and extrapolating IC projections to each electrode position into a standard projection image across the scalp. These square images, 32 pixels to a side, are calculated using a slightly modified version of the \texttt{topoplot} function in EEGLAB. 
	Power spectral densities from 1 to 100 Hz are calculated using a variation of Welch’s method \citep{pwelch} that takes the median value across time windows rather than the mean. This version was used because movement artifacts are a common occurrence in EEG datasets and the sample median is more robust to outliers than the sample mean \citep{hampel2011robust}.

	ECD model estimates are based on a three-layer boundary element method (BEM) forward-problem electrical head template (MNI) and assume that each IC scalp topography is the scalp projection of an infinitely small point-source current dipole inside the skull \citep{brazier1966study, HENDERSON1975117, adde2003symmetric}. 
	Some ICs require a dual-symmetric ECD model, likely representing the joint activation of cortical patches directly connected across the brain midline, e.g. by the corpus callosum. 
	The ECD model is fit using the DipFit plug-in in EEGLAB which calculates dipole positions and moments that best match the IC scalp topography. The better the resulting fit, the more ``dipolar" an IC can be said to be. Examples of some of these features are shown in Figure \ref{fig:website}.
	
	\subsubsection{ICLabel Website and Label Collection}
\begin{figure}
		\centering
		\includegraphics[width=\textwidth]{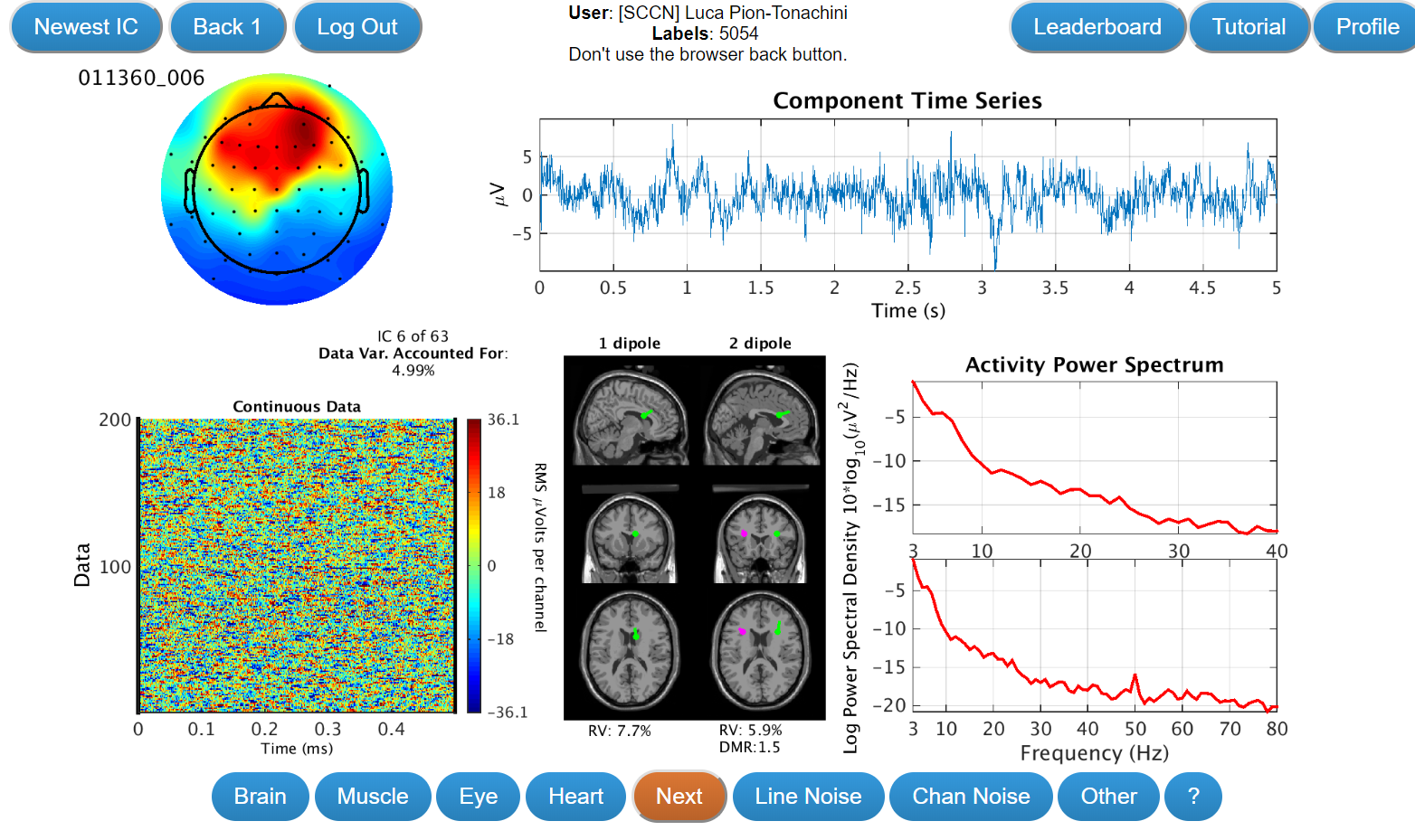}
		\caption{An IC labeling example from the ICLabel website (\url{https://iclabel.ucsd.edu/tutorial}), which also gives a detailed description of the features shown above. Label contributors are shown the illustrated IC measures and must decide which IC category or categories best apply. They mark their decision by clicking on the blue buttons below, and have the option of selecting multiple categories in the case that they cannot decide on one or believe the IC contains an additive mixture of sources. There is also a ``?" button that they can use to indicate low confidence in the submitted label.}
		\label{fig:website}
	\end{figure}

	To gather labels for ICs in the ICLabel training set, the ICLabel website (\url{https://iclabel.ucsd.edu/tutorial}) was created in the PHP scripting language using the Laravel website framework.
	With the help of over 250 contributors, henceforth referred to as ``labelers", the ICLabel website collected over 34,000 suggested labels on over 8,000 ICs through the interface illustrated in Figure \ref{fig:website}. Currently, each labeled IC has an average of 3.8 suggested labels associated with it.
	The website was advertised through the EEGLAB mailing list of EEGLAB users worldwide, and to the SCCN mailing list for lab members and visitors. The labeler pool is comprised of several IC labeling experts and many more labelers of unknown skill. To mitigate the effect of novices contributing incorrect labels to the database, the website also provides a thorough tutorial on how to recognize and label EEG ICs.
	In this way, the ICLabel website has become an educational tool. Many visitors to the website read the IC labeling tutorial and use the ``practice labeling" tool (\url{https://iclabel.ucsd.edu/labelfeedback}) that offers feedback about the labels others have assigned to the provided sample ICs. The ``practice labeling" tool currently has been used more than 49,000 times and some professors report using it to train students.
	
	\subsubsection{Crowd Labeling}
	\label{sec:cl}
	To create a coherent set of IC labels accompanying a subset of the ICs in the ICLabel training set, suggested labels collected through the ICLabel website were processed using the crowd labeling (CL) algorithm ``crowd labeling latent Dirichlet allocation" (CL-LDA) \citep{pion2017crowd}. This gave 5,937 usable labeled EEG ICs in the training set. CL algorithms estimate a single ``true label" given redundant labels for that IC provided by various labelers. This can be done multiple ways, but every CL method must reconcile disagreeing labels. CL algorithms generally do so by noting which labelers tend to agree with others and which labelers do not, upweighting and downweighting votes from those users respectively. Some methods model only the estimated labels, while others in addition model the apparent skill of each labeler; some even estimate the difficulty of the individual items being labeled.
	
	CL-LDA estimates ``true labels" as a compositional vector (vector of non-negative elements that sum to one) for each IC using the redundant labels from different labelers. Compositional labels can be thought of as softened discrete labels. In the case of ICs, this is the difference between allowing an IC to be partly ``Eye" and partly ``Muscle", or mostly ``Brain" plus some ``Line Noise", as opposed to asserting that any particular IC must be surely ``Brain" or ``Muscle" or some other class. In effect, compositional labels acknowledge that ICs may be partially ambiguous, or might not contain perfectly unmixed signals. Compositional labels can also reveal how ICs of one category may be confused with another category. Further details on CL-LDA and the specific hyperparameters used in the ICLabel dataset are given in \ref{app:cllda}.
	
	\subsection{ICLabel Expert-labeled Test Set}
	IC classification performance on the ICLabel training set is not an ideal indicator of general IC classification performance for two reasons: (1) the labels are crowdsourced, so that, even after applying CL-LDA, there are likely errors in some labels, and (2) the dataset is used many times over in the course of network and hyper-parameter optimization (described in Section \ref{sec:candidate}) which may have caused some level of implicit overfitting despite measures taken to avoid this. 
	
	For these reasons, additional datasets not present in the training set were procured and six experts were asked to label 130 ICs from those datasets. These 130 ICs comprise the ICLabel test set we used to validate the ICLabel classifier and to compare its results against existing IC classifiers. The ten additional datasets came from five different studies, two datasets from each, that had used differing recording environments, experimental paradigms, EEG amplifiers, electrode montages, preprocessing pipelines, and even ICA algorithms. These variations were purposely sought as a surrogate test of the ICLabel classifier's ability to generalize. As expert labeling is a scarce resource, only a subset of the ICs from the chosen datasets were shown to the experts for labeling. These ICs were selected by sorting the ICs within a dataset by decreasing power and taking the union among the first five ICs, five more ICs at equally spaced intervals in descending order of source power (always including the weakest IC), and the seven ICs with highest selected class probability as per the ICLabel\iclbeta{} EEGLAB plug-in for each IC category, so as to more evenly include examples of rare classes such as Heart ICs. This usually produced 12 to 13 selected ICs per dataset, giving a total of 130 ICs in the expert-labeled test set from the ten additional datasets. The six redundant expert labels per IC were also collected through the ICLabel website, a section visible only to labelers manually marked as ``experts", and were combined into a single label estimate for each IC using CL-LDA with settings detailed in \ref{app:cllda}.
	
	\subsection{ICLabel Candidate Classifiers}
	\label{sec:candidate}
	
	\begin{figure}
		\centering
		\resizebox{\textwidth}{!}{\begin{tikzpicture}[scale=1.5,font=\footnotesize,>=latex]
			\tikzset{
				box/.style={draw,thick,fill=white},
				rbox/.style={draw,thick,rounded corners=3},
				cls/.style={align=right,inner sep=1.5},
				circ/.style={draw,circle,minimum size=0.2cm},
				bg/.style={rectangle,fill=black!2,fill opacity=0.5}
			}
			
			\node[rbox,minimum width=5cm,minimum height=5cm,fill=blue!2,label={[anchor=north west]north west:Discriminator/Classifier Network}](D) at (2.5,0) {};
			\node[box,minimum width=2.15cm,minimum height=1cm](Dtopo) at (1.95,1) {Topographies};
			\node[box,minimum width=2.15cm,minimum height=1cm](Dpsd) at (1.95,0) {PSD};
			\node[box,minimum width=2.15cm,minimum height=1cm](Dac) at (1.95,-1) {Autocorrelation};
			\node[box,minimum width=0.5cm,minimum height=3cm](Dstack) at (3.1,0) {};
			\draw[->] (D.west) -- (Dtopo.west);
			\draw[->] (D.west) -- (Dpsd.west);
			\draw[->] (D.west) -- (Dac.west);
			\draw[->] (Dtopo.east) -- (Dstack);
			\draw[->] (Dpsd.east) -- (Dstack);
			\draw[->] (Dac.east) -- (Dstack);
			\draw[->] (Dstack) -- ($(Dstack) + (1.5, 0)$) node[pos=0.33,above]{Output};
			
			\scoped[xshift=-5cm,yshift=-1.3cm]{
				\node[rbox,minimum width=6cm,minimum height=4.4cm,fill=green!2,label={[anchor=north west]north west:Actual ICs}](R) at (2,-0) {};
				\node[rbox,minimum width=5cm,minimum height=1.8cm,label={[anchor=north west]north west:Unlabeled}](Ru) at (2,0.6) {};
				\node[rbox,minimum width=5cm,minimum height=1.8cm,label={[anchor=north west]north west:Labeled}](Rl) at (2,-0.8) {};
				
				\node[inner sep=0pt, left=1.5 of Rl.center, anchor=center](Rtopoex3) {\includegraphics[scale=0.1]{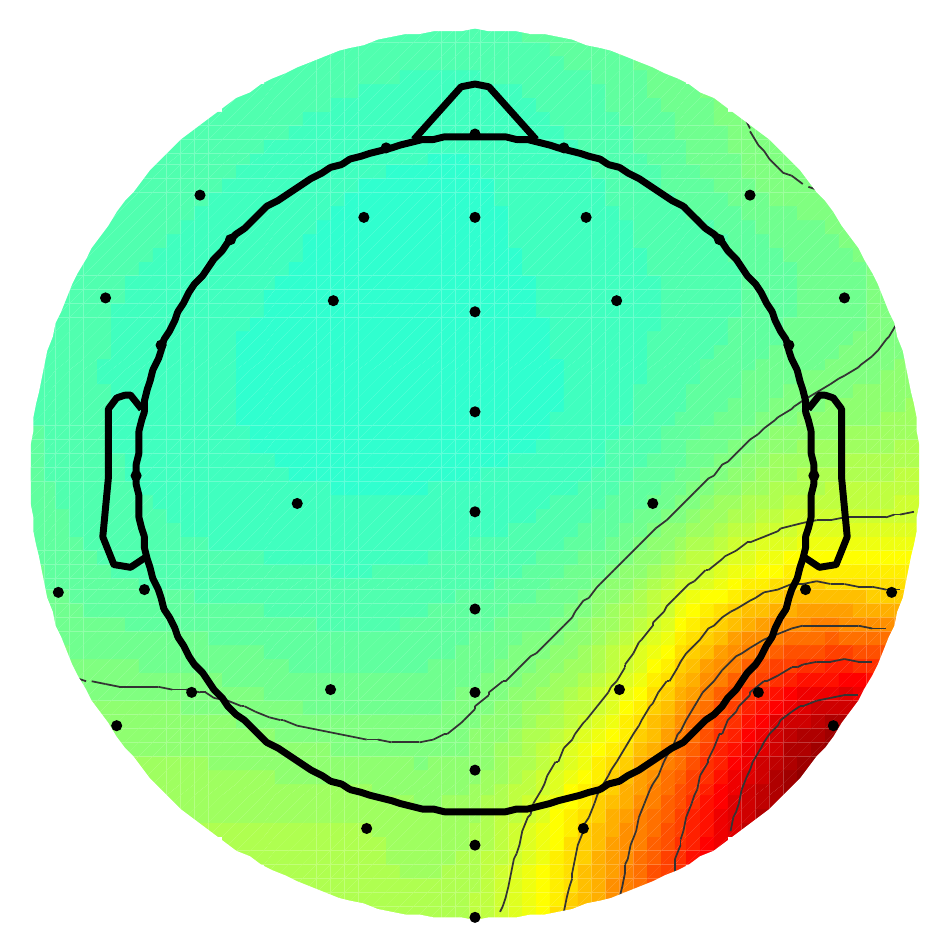}};
				\node[inner sep=0pt,below left=0.2 of Rtopoex3.north east](Rtopoex2) {\includegraphics[scale=0.1]{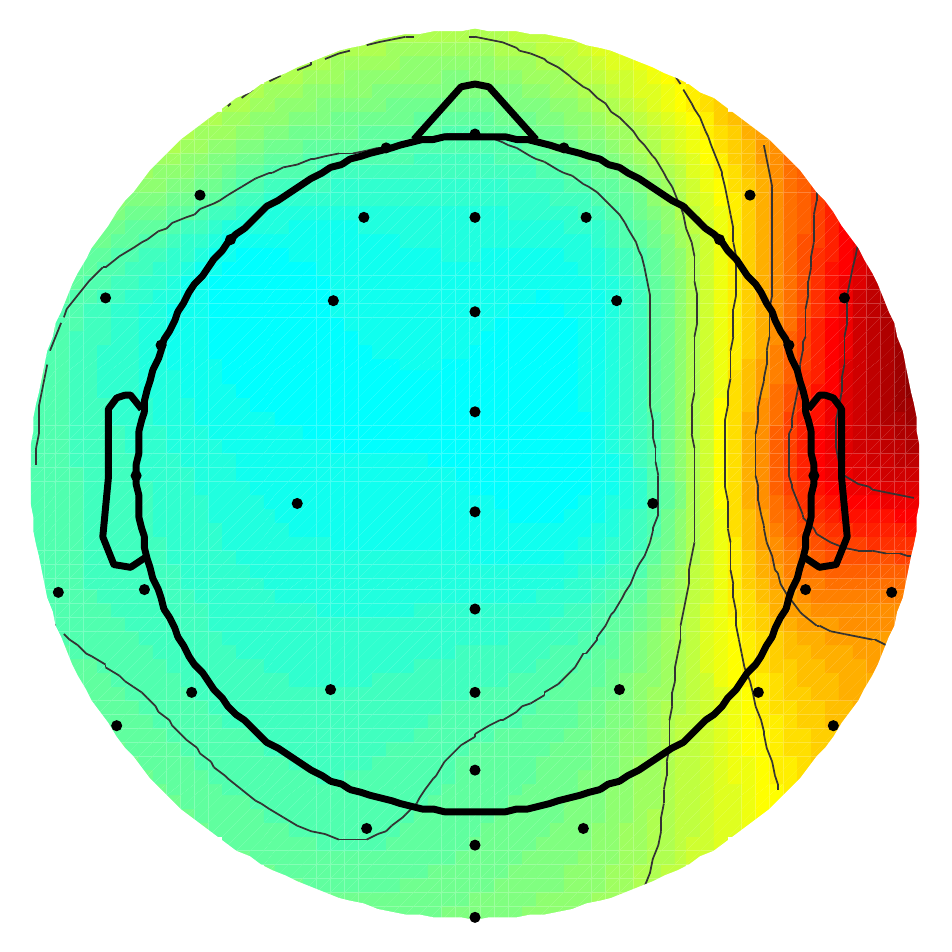}};
				\node[inner sep=0pt,below left=0.2 of Rtopoex2.north east](Rtopoex1) {\includegraphics[scale=0.1]{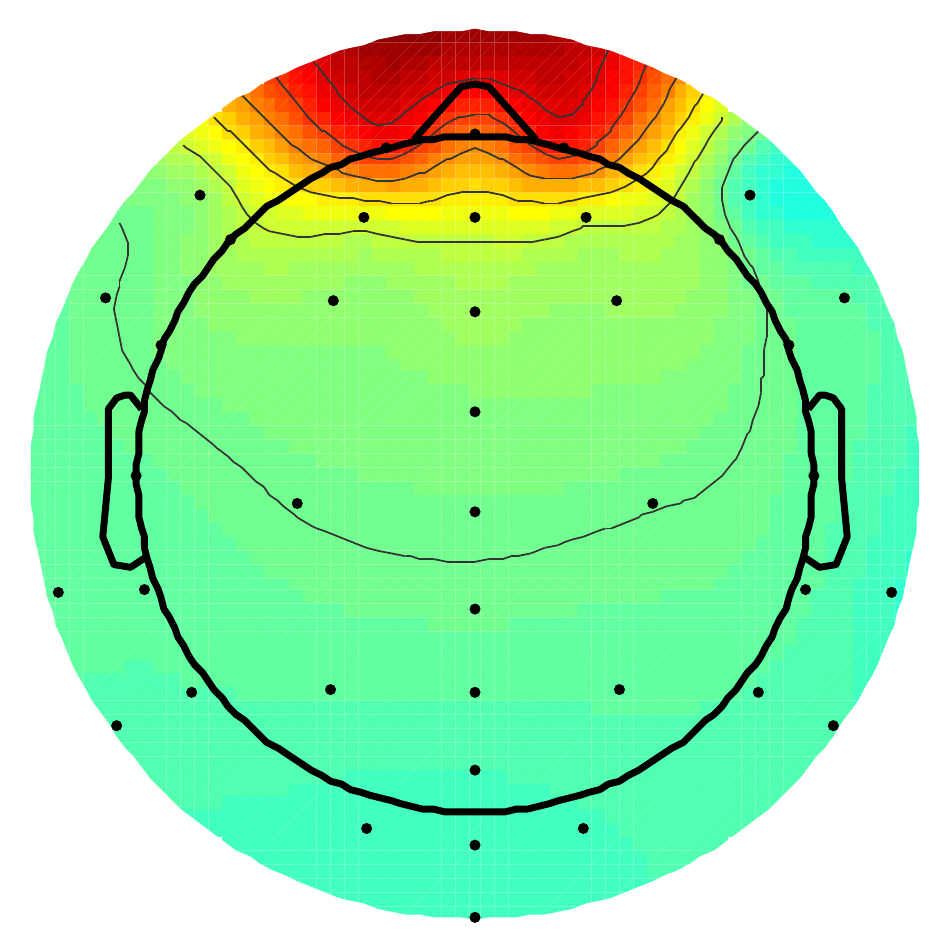}};
				
				\node[bg,inner sep=0pt, right=0.2 of Rl.center, anchor=center](Rpsdex3) {\includegraphics[scale=0.1]{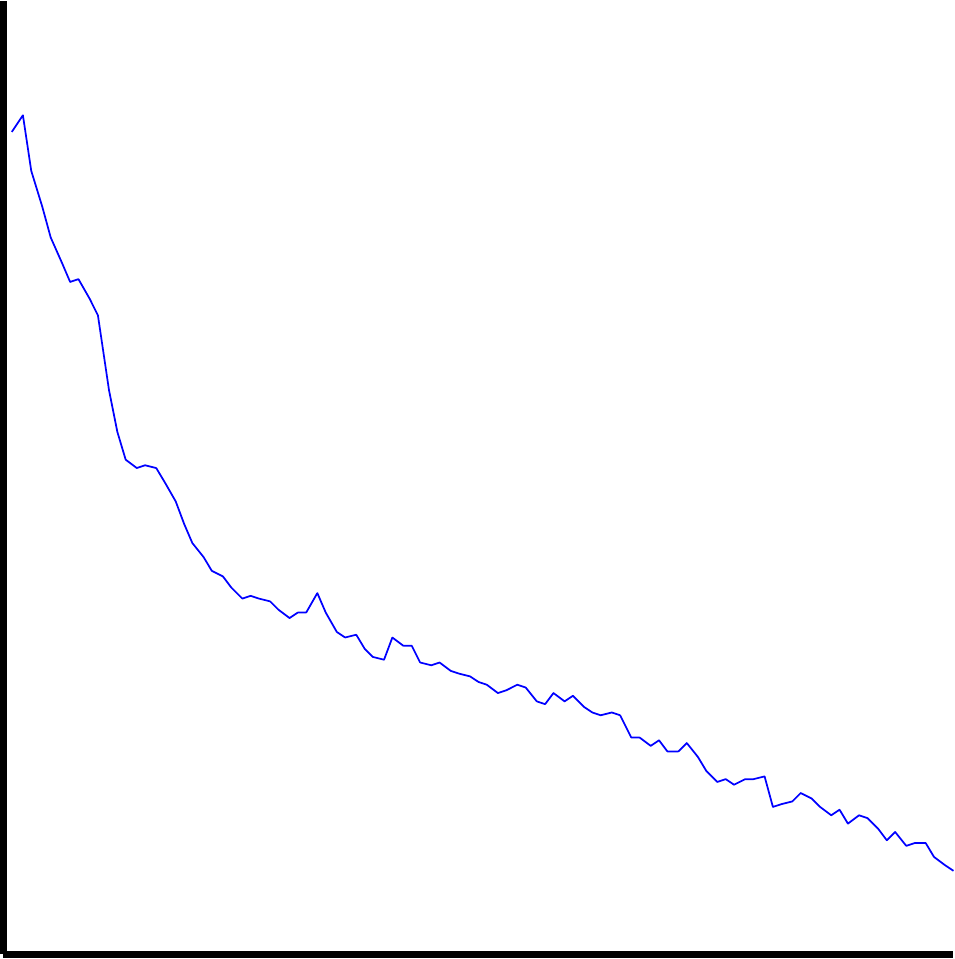}};
				\node[bg,inner sep=0pt,below left=0.2 of Rpsdex3.north east](Rpsdex2) {\includegraphics[scale=0.1]{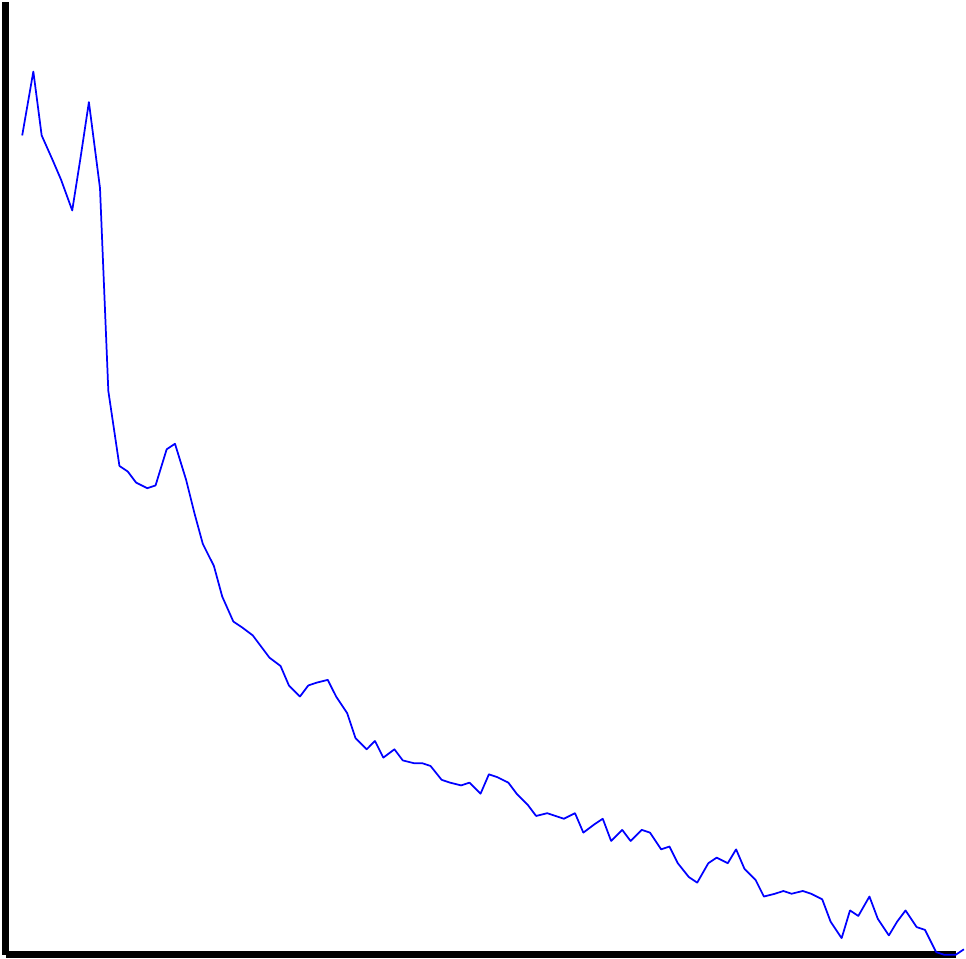}};
				\node[bg,inner sep=0pt,below left=0.2 of Rpsdex2.north east](Rpsdex1) {\includegraphics[scale=0.1]{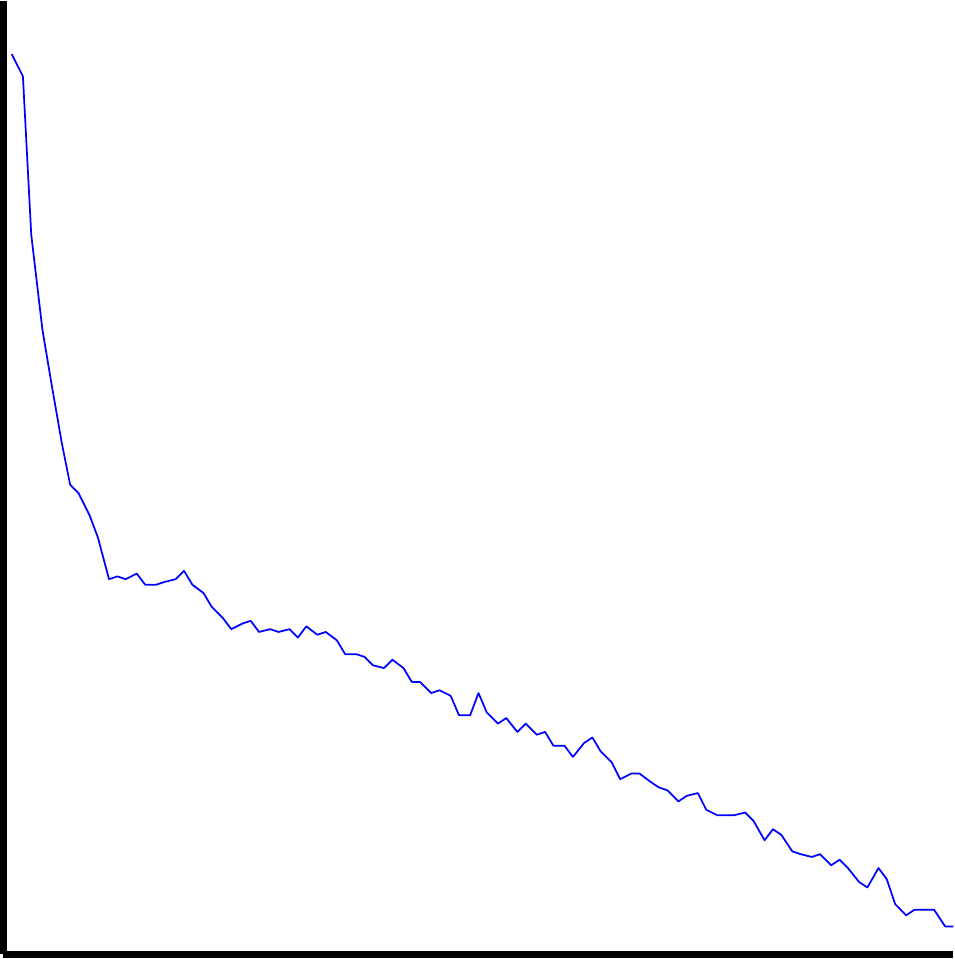}};
				
				\node[bg,inner sep=0pt,right=1.9 of Rl.center, anchor=center](Racex3) {\includegraphics[scale=0.1]{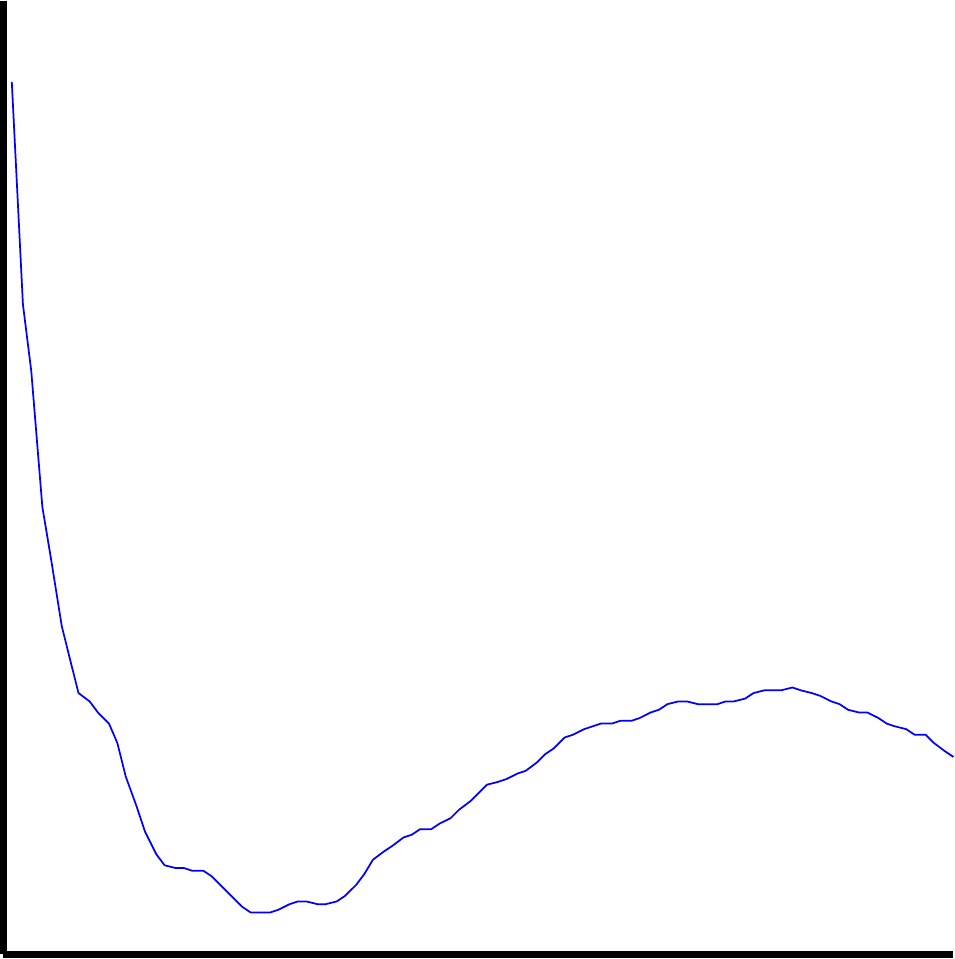}};
				\node[bg,inner sep=0pt,below left=0.2 of Racex3.north east](Racex2) {\includegraphics[scale=0.1]{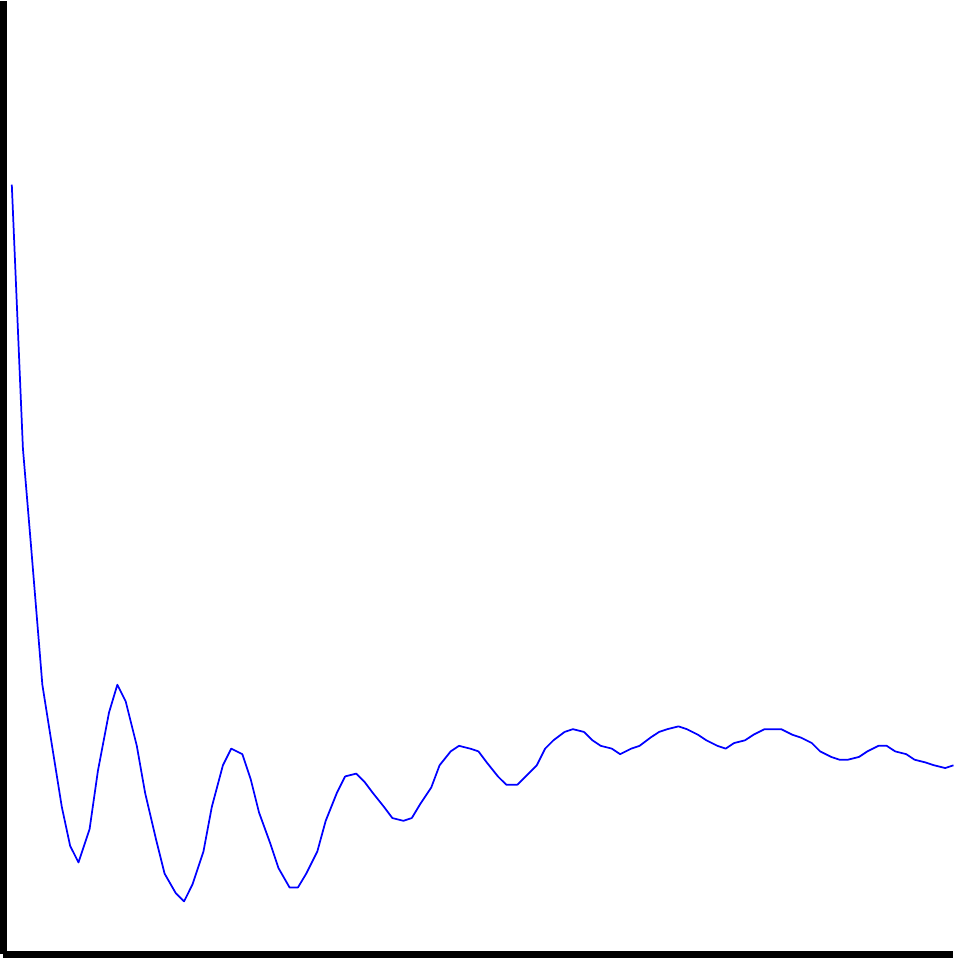}};
				\node[bg,inner sep=0pt,below left=0.2 of Racex2.north east](Racex1) {\includegraphics[scale=0.1]{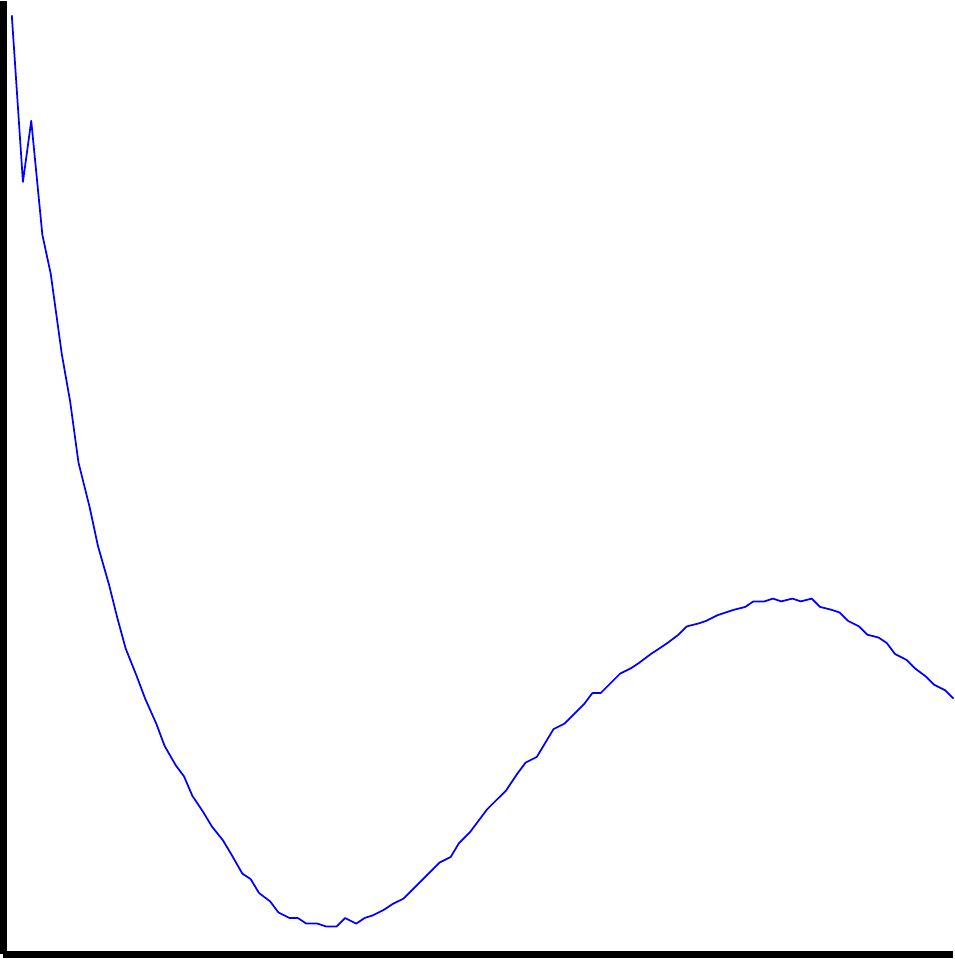}};
				
				\node[inner sep=0pt, left=1.5 of Ru.center, anchor=center](Rtopoex3) {\includegraphics[scale=0.1]{topo3}};
				\node[inner sep=0pt,below left=0.2 of Rtopoex3.north east](Rtopoex2) {\includegraphics[scale=0.1]{topo2}};
				\node[inner sep=0pt,below left=0.2 of Rtopoex2.north east](Rtopoex1) {\includegraphics[scale=0.1]{topo1}};
				
				\node[bg,inner sep=0pt, right=0.2 of Ru.center, anchor=center](Rpsdex3) {\includegraphics[scale=0.1]{psd3}};
				\node[bg,inner sep=0pt,below left=0.2 of Rpsdex3.north east](Rpsdex2) {\includegraphics[scale=0.1]{psd2}};
				\node[bg,inner sep=0pt,below left=0.2 of Rpsdex2.north east](Rpsdex1) {\includegraphics[scale=0.1]{psd1}};
				
				\node[bg,inner sep=0pt, right=1.9 of Ru.center, anchor=center](Racex3) {\includegraphics[scale=0.1]{ac3}};
				\node[bg,inner sep=0pt,below left=0.2 of Racex3.north east](Racex2) {\includegraphics[scale=0.1]{ac2}};
				\node[bg,inner sep=0pt,below left=0.2 of Racex2.north east](Racex1) {\includegraphics[scale=0.1]{ac1}};
			}
			
			\scoped[xshift=-5cm,yshift=2cm]{
				\node[box,dashed,color=teal,minimum width=8cm,minimum height=8.1cm,fill=none,  label={[anchor=north west,text=teal]north west:Semi-Supervised}](SS) at (2.3,-0.7) {};
				\node[rbox,minimum width=6cm,minimum height=5cm,fill=red!2,label={[anchor=north west]north west:Generator Network}](G) at (2,0) {};
				\node[box,minimum width=2.15cm,minimum height=1cm](Gtopo) at (1.8,1) {Topographies};
				\node[box,minimum width=2.15cm,minimum height=1cm](Gpsd) at (1.8,0) {PSD};
				\node[box,minimum width=2.15cm,minimum height=1cm](Gac) at (1.8,-1) {Autocorrelation};
				\node[box,minimum height=1.5cm, label=Noise, left=of Gpsd](Gnoise) {};
				\pgfmathsetseed{1234}
				\pgfmathsetmacro\height{1} 
				\pgfmathsetmacro\n{100}
				\foreach \p in {1,...,\n} {
					\pgfmathsetmacro\flag{rand>0 ? 0:1}
					\ifnum1=\flag
					\pgfmathsetmacro\topBorder{\height * (\p - 1) / \n]]}
					\pgfmathsetmacro\bottomBorder{\height * \p / \n}
					\fill[opacity=0.5] ($(Gnoise.north west) - (0,\topBorder)$) rectangle ($(Gnoise.north east) - (0,\bottomBorder)$);
					\fi
				}
				\draw[->] (Gnoise) -- (Gtopo.west);
				\draw[->] (Gnoise) -- (Gpsd.west);
				\draw[->] (Gnoise) -- (Gac.west);
				
				\node[inner sep=0pt,above right=-0.2 and 2 of Gtopo.center](Gtopoex3) {\includegraphics[scale=0.1]{topo3}};
				\node[inner sep=0pt,below left=0.2 of Gtopoex3.north east](Gtopoex2) {\includegraphics[scale=0.1]{topo2}};
				\node[inner sep=0pt,below left=0.2 of Gtopoex2.north east](Gtopoex1) {\includegraphics[scale=0.1]{topo1}};
				
				\node[bg,inner sep=0pt,above right=-0.2 and 2 of Gpsd.center](Gpsdex3) {\includegraphics[scale=0.1]{psd3}};
				\node[bg,inner sep=0pt,below left=0.2 of Gpsdex3.north east](Gpsdex2) {\includegraphics[scale=0.1]{psd2}};
				\node[bg,inner sep=0pt,below left=0.2 of Gpsdex2.north east](Gpsdex1) {\includegraphics[scale=0.1]{psd1}};
				
				\node[bg,inner sep=0pt,above right=-0.2 and 2 of Gac.center](Gacex3) {\includegraphics[scale=0.1]{ac3}};
				\node[bg,inner sep=0pt,below left=0.2 of Gacex3.north east](Gacex2) {\includegraphics[scale=0.1]{ac2}};
				\node[bg,inner sep=0pt,below left=0.2 of Gacex2.north east](Gacex1) {\includegraphics[scale=0.1]{ac1}};
				
				\draw[->] (Gtopo) -- (Gtopoex1);
				\draw[->] (Gpsd) -- (Gpsdex1);
				\draw[->] (Gac) -- (Gacex1);
			}
			
			\node[circ,left=1.3 of Dpsd](Dnode){};
			\node[circ](Gnode) at ($(Dnode|-G) - (1,0)$) {};
			\node[circ](Runode) at (Gnode|-Ru.center){};
			\node[circ](Rlnode) at (Runode|-Rl.center){};
			
			\draw (Dnode) -- (D);
			\draw[->] (G) -- (Gnode);
			\draw[->] (Ru) -- (Runode);
			\draw[->] (Rl) -- (Rlnode);
			\draw[->] (Gnode.east) -- (Dnode);
			\draw[->] (Runode.east) -- (Dnode);
			\draw[->] (Rlnode.east) -- (Dnode);
			\end{tikzpicture}
		}
		\caption{Candidate artificial neural network (ANN) architectures tested in developing the ICLabel classifier. White rectangles represent ANN blocks comprised of one or more convolutional layers; arrows indicate information flow. The section in the upper left labeled ``Semi-Supervised" (teal dashed outline) was only present in the GAN paradigm during training and was used to generate simulated IC features to compare against unlabeled training examples from the ICLabel training set. The box to the right labeled ``Discriminator" remained nearly identical in structure for all three training paradigms (although the parameters used in the final learned network differed). Convergence of arrows into the classifier network indicates the input sources for the classifier during training and does \emph{not} imply data combination, e.g. through summation. After training is complete, classifiers were given \emph{unlabeled} ICs to classify. See \ref{app:arch} for a detailed description of the ANN implementations.}
		\label{fig:arch}
	\end{figure}
	
	Multiple candidate classifiers were trained and compared to select the architecture and training paradigm best suited for creating the final ICLabel classifier. These candidate versions differed in the feature sets used as inputs, in training paradigm, and in model structure. In this way the ICLabel training set was used to train six candidate ICLabel classifiers. Three artificial neural network (ANN) architectures were tested; all had the same underlying convolutional neural network (CNN) structure used for inference. Figure \ref{fig:arch} graphically summarizes the three ANN architectures of the ICLabel candidates. Two of those architectures were CNNs trained on only the labeled ICs. The first of those CNNs optimized an unweighted cross entropy loss while the second optimized a weighted cross entropy loss that doubly weighted Brain IC classification errors (wCNN). Cross entropy is a mathematical function that compares two class probability vectors (typically label vectors) and produces a scalar output related to how similar those two vector are. See \ref{app:metrics} for a more detailed explanation. The third classifier architecture was based on a variation of semi-supervised learning generative adversarial networks (SSGAN) \citep{odena2016semi,salimans2016improved}, an extension of generative adversarial networks (GAN) \citep{goodfellow2014generative}. Detailed descriptions of the ICLabel candidate classifier inputs, architectures, and training paradigms are given in \ref{app:arch} for the two CNNs and \ref{app:gan} for the GAN. 
	
	Each of the three network architectures described here were further differentiated by associating them with two possible groups of input feature sets. The first group used scalp topographies and PSDs as inputs, while the second group also used autocorrelation functions. The other feature sets included in the full ICLabel training set were not used by the candidate classifiers as they were either too computationally expensive to compute or were found to not contribute new information in preliminary evaluations beyond the information provided by the scalp topographies, PSDs, and autocorrelation functions.
	
	As described in \ref{app:arch}, the ICLabel training set was augmented to four times its original size by exploiting left--right and positive--negative symmetries in scalp topographies. This augmentation was not repeated for the expert-labeled test set. Instead, the final ICLabel classifier internally duplicates each IC to exploit the two scalp topography symmetries and takes the average of the four resulting classifications. 
	
	\subsection{Evaluation} \label{sec:eval}
	To select the candidate classifier that would become the released ICLabel classifier, six candidate versions of the ICLabel classifier were tested using a three-by-two factorial design with repeated measures on the ICLabel training set. The first factor, ANN architecture, had three levels (described in Section \ref{sec:candidate}): (1) GAN, (2) CNN, and (3) wCNN. The second factor, feature sets provided to the classifiers, had two levels: (1) networks using only scalp topographies and PSDs and (2) networks also using autocorrelation functions. Below, use of the autocorrelation feature set is indicated by a subscript ``AC" following the architecture, as in GAN\ac{}.
	
	To compare the performance of candidate classifiers, the labeled portion of the ICLabel training set was split so as to follow a ten-fold stratified cross-validation scheme. Within each fold, the data were split into training, validation, and testing data (at a ratio of 8:1:1) in a way that attempted to maintain equal class proportions across the three subsets of the labeled data. The training data from each fold was used to train every candidate classifier version, and that fold's validation data was used to determine when to stop training with early stopping \citep{Prechelt2012}. Each fold's test data was used to calculate the performance of all classifiers trained on that fold's training data. Overall performance for each candidate classifier was taken as the average performance measured across all ten folds. While not relevant to candidate classifier selection, performance of some published IC classification methods was also calculated on the same cross-validation folds. To not waste any training data, the training paradigm that produced the best performing ICLabel candidate was then used to train a new classifier using the best performing candidate architecture with the \emph{entire} ICLabel training set, minus 400 labeled examples now held out as a validation set for early stopping. The resulting classifier became the official ICLabel classifier and was compared to existing methods on the expert-labeled test set.
	
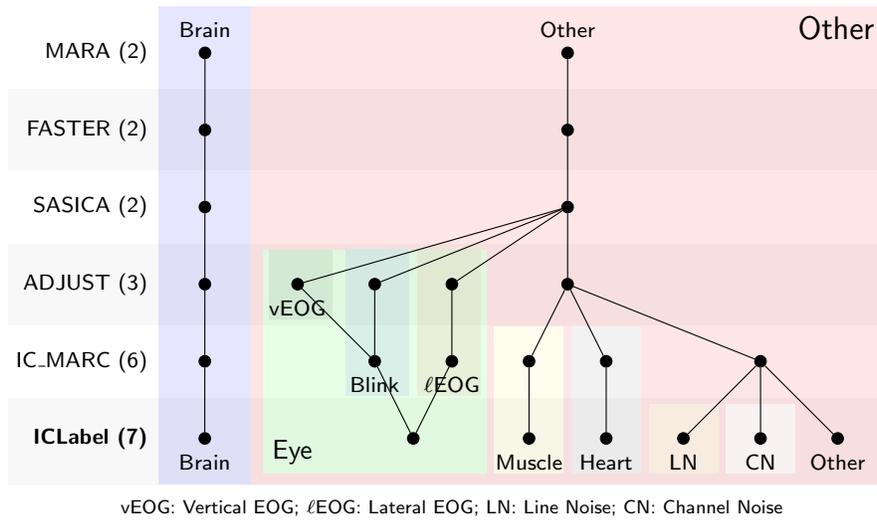
\begin{figure}
		\centering
		\resizebox{\textwidth}{!}{
			\begin{tikzpicture}[scale=1.5,font=\footnotesize]
			\tikzset{
				solid node/.style={circle,draw,inner sep=1.5,fill=black,on grid},
				hollow node/.style={circle,draw,inner sep=1.5,on grid},
				cls/.style={align=right,inner sep=1.5},
			}
			
			\node(other0)[solid node, label={Other}]{};
			\node(other1)[solid node, below=of other0]{};
			\node(other2)[solid node, below=of other1]{};
			
			\node(verteye0)[solid node, below left=1 and 3.5 of other2, label={270:vEOG}]{};
			\node(blink0)[solid node, below left=1 and 2.5 of other2]{};
			\node(lateye0)[solid node, below left=1 and 1.5 of other2]{};
			\node(blink1)[solid node, below=of blink0, label={270:Blink}]{};
			\node(lateye1)[solid node, below=of lateye0, label={270:$\ell$EOG}]{};
			\node(eye1)[solid node, below right=1 and 0.5 of blink1]{};
			
			\node(other3)[solid node, below=of other2]{};
			\node(muscle0)[solid node, below left=1 and 0.5 of other3]{};
			\node(muscle1)[solid node, below=of muscle0, label={270:Muscle}]{};
			\node(heart0)[solid node, below right=1 and 0.5 of other3]{};
			\node(heart1)[solid node, below=of heart0, label={270:Heart}]{};
			
			\node(other4)[solid node, below right=1 and 2.5 of other3]{};
			\node(linen)[solid node, below left=1 and 1 of other4, label={270:LN}]{};
			\node(chann)[solid node, below=of other4, label={270:CN}]{};
			\node(other5)[solid node, below right=of other4, label={270:Other}]{};
			
			\draw[-] (other0) to (other1);
			\draw[-] (other1) to (other2);
			\draw[-] (other2) to (other3);
			\draw[-] (other3) to (other4);
			\draw[-] (other4) to (other5);
			
			\draw[-] (other2) to (lateye0);
			\draw[-] (other2) to (blink0);
			\draw[-] (other2) to (verteye0);
			\draw[-] (lateye0) to (lateye1);
			\draw[-] (blink0) to (blink1);
			\draw[-] (verteye0) to (blink1);
			\draw[-] (lateye1) to (eye1);
			\draw[-] (blink1) to (eye1);
			
			\draw[-] (other3) to (muscle0);
			\draw[-] (muscle0) to (muscle1);
			\draw[-] (other3) to (heart0);
			\draw[-] (heart0) to (heart1);
			
			\draw[-] (other4) to (linen);
			\draw[-] (other4) to (chann);
			
			\node(brain0)[solid node, left=4.7 of other0, label=Brain]{};
			\node(brain1)[solid node, below=of brain0]{};
			\node(brain2)[solid node, below=of brain1]{};
			\node(brain3)[solid node, below=of brain2]{};
			\node(brain4)[solid node, below=of brain3]{};
			\node(brain5)[solid node, below=of brain4, label={270:Brain}]{};
			\draw[-] (brain0) to (brain1);
			\draw[-] (brain1) to (brain2);
			\draw[-] (brain2) to (brain3);
			\draw[-] (brain3) to (brain4);
			\draw[-] (brain4) to (brain5);
			
			\node(mara)[cls,left=0.6 of brain0]{MARA (2)};
			\node(faster)[cls,left=0.6 of brain1]{FASTER (2)};
			\node(sasica)[cls,left=0.6 of brain2]{SASICA (2)};
			\node(adjust)[cls,left=0.6 of brain3]{ADJUST (3)};
			\node(icmarc)[cls,left=0.6 of brain4]{IC\_MARC (6)};
			\node(icl)[cls,left=0.6 of brain5]{\textbf{ICLabel (7)}};
			
			\def \sss {0.08}
			\def \ss {0.25}
			\def \ms {0.3}
			\def \mms {0.35}
			\def \ls {0.4}
			\def \vls {1.7}
			
			\scoped[on background layer]{
%
				\fill [blue!10] ($(brain0) + (-\ls, \ls)$) rectangle ($(brain5) + (\ls, -\ls)$);
				
				\fill [red!10] ($(other5|-other0) + (+\ls, \ls)$) rectangle ($(verteye0|-other5) + (-\ls, -\ls)$) node[black,pos=0,below left] {\large Other};
				\fill [green!10] ($(lateye1|-verteye0) + (\ms, \ms)$) rectangle ($(verteye0|-eye1) + (-\ms, -\ms)$) node[black,pos=1,above right] {\normalsize Eye};
				\fill [black!40!green!15] ($(verteye0) + (-\ss, \ms)$) rectangle ($(verteye0) + (\ms, -\ms)$);
				\fill [blue!40!green!15] ($(blink0) + (-\ss, \ms)$) rectangle ($(blink1) + (\ms, -\ms)$);
				\fill [red!40!green!15] ($(lateye0) + (-\ms, \ms)$) rectangle ($(lateye1) + (\ss, -\ms)$);
				\fill [yellow!8] ($(muscle0) + (-\ms, \ms)$) rectangle ($(muscle1) + (\ms, -\ms)$);
				\fill [black!5] ($(heart0) + (-\ms, \ms)$) rectangle ($(heart1) + (\ms, -\ms)$);
				\fill [orange!10] ($(linen) + (\ms, \ms)$) rectangle ($(linen) + (-\ms, -\ms)$);
				\fill [pink!10] ($(chann) + (\ms, \ms)$) rectangle ($(chann) + (-\ms, -\ms)$);
				
				\fill [black,opacity=0.03] ($(brain5) + (-\vls, \mms)$) rectangle ($(other5) + (\ls, -\ls)$) node[pos=0](tl){} node[pos=1](br){};
				\fill [black,opacity=0.03] ($(brain3) + (-\vls, \mms)$) rectangle ($(other5|-brain3) + (\ls, -\mms)$);
				\fill [black,opacity=0.03] ($(brain1) + (-\vls, \mms)$) rectangle ($(other5|-brain1) + (\ls, -\mms)$);
			}
			
			\node(abbreviations) at ($(other0|-icl) + (-1, -0.6)$]) {\scriptsize vEOG: Vertical EOG; $\ell$EOG: Lateral EOG; LN: Line Noise; CN: Channel Noise};
			\end{tikzpicture}
		}
		\caption{Categories labeled by the IC classifiers that were evaluated on the expert-labeled test set. The top five classifiers listed on the vertical axis are described in Section \ref{sec:prior_methods}. The tree structure and colored boxes connecting labels of different classifiers signifies how the classifier labels are related and how they could be merged to allow comparisons between classifiers with non-identical IC categories. For example, all IC classifiers can be compared across two classes by merging all categories contained within the red box into the overarching category of Other ICs. Similarly, all categories in the green box can be simplified to form a single Eye IC category. The following acronyms are used in the above figure: ``vEOG" for ``vertical EOG activity", ``$\ell$EOG" for ``lateral EOG activity", ``LN" for ``Line Noise", and ``CN" for ``Channel Noise".}
		\label{fig:classes}
	\end{figure}

	Performance comparisons between the candidate IC classifiers required a fixed set of IC classes over which to compare scores. As most IC classifiers discriminate between differing sets of IC categories, both in number and interpretation, it was necessary to merge label categories to allow direct classifier comparisons. At one extreme, IC labels and predictions can be reduced to either ``Brain" or ``Other" to allow comparison of nearly all the IC classifiers. Further subsets could be used for three-, five- and seven-class comparisons, as detailed in Figure \ref{fig:classes}. This study used the five-class and seven-class comparisons as well as the already-described two-class comparison. The five-class comparison combined all eye-related IC categories into a unified Eye IC category and all non-biological artifact ICs and unknown-source ICs into a unified Other IC category. The five-class comparison allowed comparison between the ICLabel candidates and final classifier and all IC\_MARC versions, while the seven-class case only allowed comparisons between ICLabel candidates and final classifier.
	
	Classifier performance was measured by comparing balanced accuracy and normalized confusion matrices after discretizing IC labels and predictions, receiver operating characteristic (ROC) curves after discretizing IC labels, ROC equivalent measures from ``soft" confusion matrices \citep{BELEITES201312} termed here as \emph{soft operating characteristics} (SOC) points, cross-entropy, and required time to calculate the IC classifications. Further explanation of these measures is given in \ref{app:metrics}.

	\section{Results}
	
	\subsection{ICLabel and Prior Methods} 
	\label{sec:result-test}
	
	\begin{table}
		\centering
		\begin{tabular}{llcc}
			\toprule
			Classes & Classifier & Balanced Accuracy & Cross Entropy\\
			& & $\frac{1}{C} \sum_{i=1}^{C} \frac{\text{TP}_i}{\text{TP}_i + \text{FN}_i}$ & $\sum_i t_i \log p_i$ \\
			\midrule
			2 & ICLabel\lite{} & \bfseries{0.855} & \bfseries{0.339} \\
			& ICLabel & \bfseries{0.841} & \bfseries{0.342} \\
			& \icmef{} & 0.816 & 0.977 \\
			& \icmsf{} & \bfseries{0.870} & \bfseries{0.377} \\
			& ADJUST & 0.585 & - \\
			& MARA & 0.757 & 0.730 \\
			& FASTER & 0.578 & - \\
			& SASICA & 0.775 & - \\
			\midrule
			5 & ICLabel\lite{} & \bfseries{0.623} & \bfseries{0.938} \\
			& ICLabel & \bfseries{0.613} & \bfseries{0.924} \\
			& \icmef{} & 0.532 & 2.659 \\
			& \icmsf{} & 0.578 & 0.982 \\
			\midrule
			7 & ICLabel\lite{} & 0.579 & 1.287 \\
			& ICLabel & 0.597 & 1.251 \\
			\bottomrule
			&&&\\
		\end{tabular}
		\caption{Scalar performance measures of the tested publicly available independent component (IC) classifiers for different numbers of IC categories. Higher balanced accuracy and lower cross entropy indicate better classification performance.}
		\label{table:scores}
	\end{table}
	
\begin{figure}
		\centering
		\includegraphics[width=0.85\textwidth]{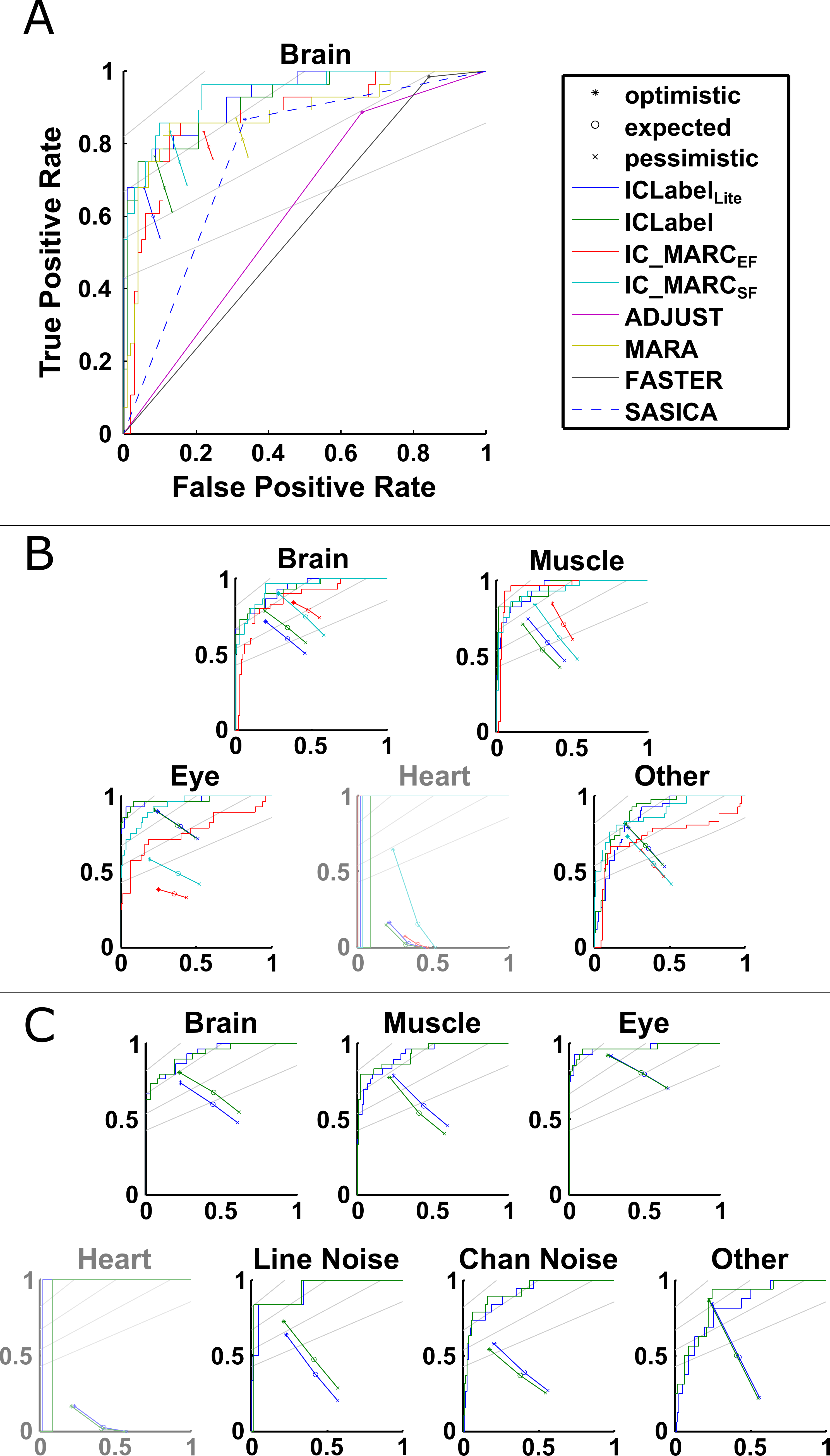}
		\caption{
			Comparison of ICLabel classification performance to that of several alternative publicly available IC classifiers. ROC curves and soft operating characteristics (SOC) points for the (A) two-class, (B) five-class, and (C) seven-class performances on the expert-labeled test set. Gray lines indicate F\textsubscript{1} score isometrics of 0.9, 0.8, 0.7, and 0.6 (from top to bottom). ``Heart" plots have been grayed out because experts marked only one IC as being heart-related leading to largely uninformative SOC points and ROC curves for that category. Refer to \ref{app:metrics} for definitions of F\textsubscript{1} score, ROC curves (traced out by the detection threshold parameter), and SOC points (shown for optimistic, expected, and pessimistic performance estimates as described in Appendix A).}
		\label{fig:roc}
	\end{figure}

	The ICLabel classifier and the ICLabel\lite{} classifier, created as described at the end of \ref{app:candidate_eval}, were compared against previously-existing, publicly-available IC classifiers.
	As described in Section \ref{sec:eval}, all IC categories besides ``Brain" must be conflated to allow a comparison across all IC classification methods simultaneously on the expert-labeled test set.
	Considering balanced accuracy (higher values are better) and cross entropy (lower values are better) as shown in Table \ref{table:scores}, in addition to ROC curves for the two-class case as shown in Figure \ref{fig:roc}, the only previously existing classifier competitive with ICLabel was \icmsf{}. IC\_MARC and ICLabel classifiers can be meaningfully compared across five IC categories, as shown in Figure \ref{fig:classes}, and disregarding the other classifiers eliminates the need to aggressively merge non-Brain ICs, allowing a more detailed comparison.
	
	
\begin{figure}
		\centering
		\includegraphics[width=\textwidth]{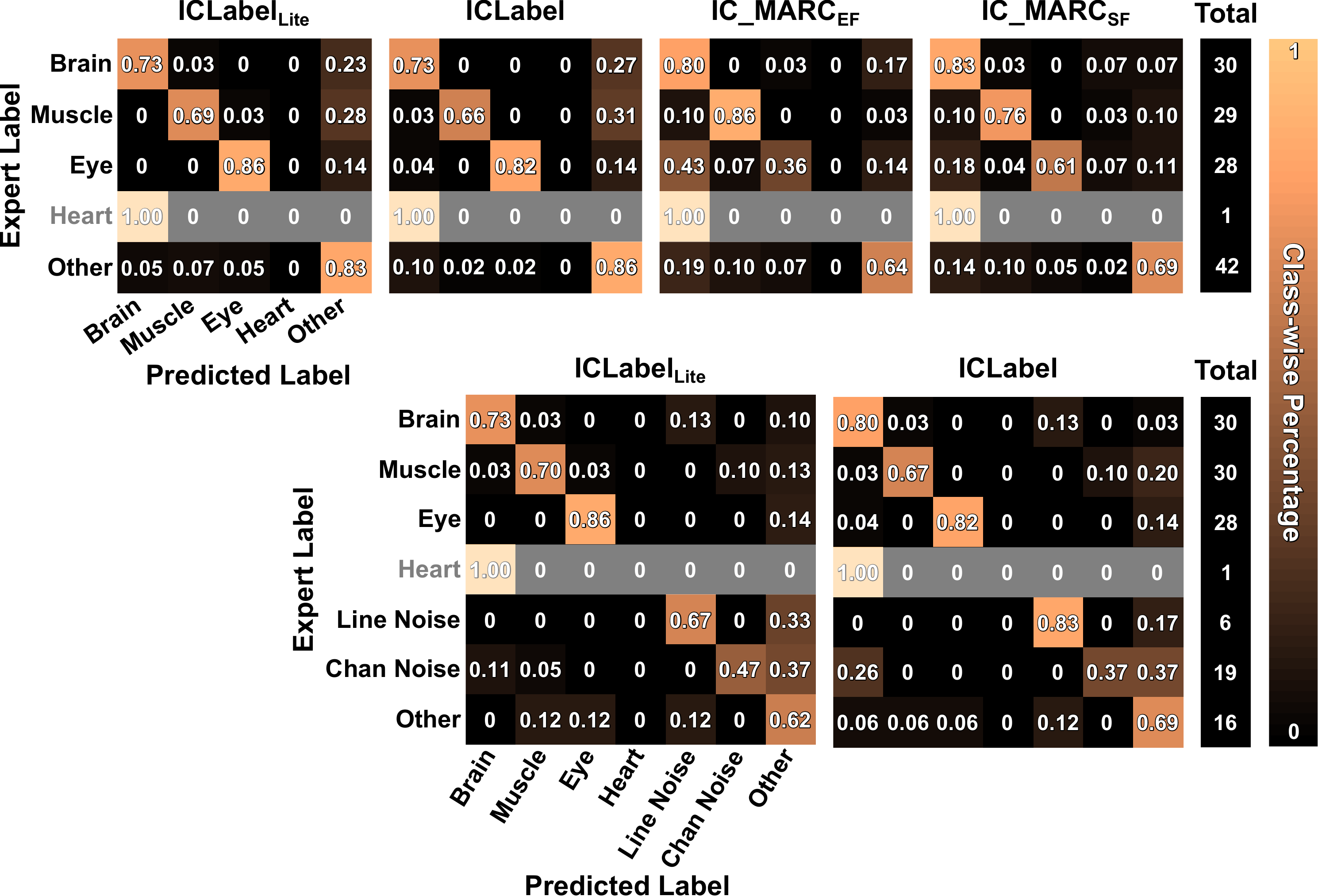}
		\caption{Normalized ICLabel and IC\_MARC confusion matrices calculated from the expert-labeled test set using five classes (top row) and seven classes (bottom row). Rows and columns of each confusion matrix contain all ICs labeled as a particular class by experts and the classifiers, respectively. Rows were normalized to sum to one such that each element along the diagonal represents the true-positive-rate (recall) for that IC category. The ``Total" columns on the right indicate how many ICs were labeled as each class by the experts (used for normalization). ``Heart" rows have been grayed out because experts marked only one IC as being heart-related leading to largely uninformative results for that row.}
		\label{fig:cmat}
	\end{figure}

	In the five-class comparison, \icmsf{} showed marginally better performance than ICLabel when classifying Brain ICs, as measured by ROC curves. SOC points indicated comparable performance whereby \icmsf{} achieved a slightly higher soft-TPR than ICLabel at the cost of also having higher soft-FPR.
	For Muscle ICs, \icmef{} outperformed all other methods as per the ROC curves, despite underperforming on nearly every other measure. Among the three other methods, \icmsf{} achieved a higher recall for Muscle ICs after thresholding labels and predictions, as seen in the second row of each five-class confusion matrix (top row of Figure \ref{fig:cmat}), despite the corresponding ROC curve not being superior to those of either ICLabel method. Both ICLabel methods performed exceptionally well on Eye ICs, greatly outperforming both IC\_MARC versions, as indicated by both the SOC points and ROC curves.
	
	Even though results are shown for Heart ICs, the expert labelers only communally selected one IC as ``Heart" and, therefore, the statistical power of results regarding Heart ICs is too low to warrant further discussion.
	With regard to Other ICs, ICLabel and ICLabel\lite{} directly outperformed both IC\_MARC models as measured by SOC points while ICLabel and \icmsf{} shared the best performance in different regimes of the performance plane as shown by their respective ROC curves.
	The confusion matrices of Figure \ref{fig:cmat} indicate that most ICLabel errors were derived from over-classifying ICs as ``Other", while the causes of \icmsf{} errors are difficult to infer.
	
	ICLabel and ICLabel\lite{} ROC curves remained nearly unchanged in the seven-class case compared to the five-class case except for Other ICs. SOC points gave similar results, although the distance between optimistic, expected, and pessimistic estimates are larger due to the increased number of IC categories. 
	The additional Line Noise IC and Channel Noise IC categories were classified relatively well, as indicated by the ROC curves, although the scarcity of Line Noise ICs in the expert-labeled test set produced low-resolution ROC curves.
	SOC points indicate some level of disagreement between the experts and ICLabel with regards to the overall label composition on these two IC categories due to the lower soft TPR values shown.
	The seven-class confusion matrix showed ICLabel to have much lower accuracy on Channel Noise ICs than would be expected from the ROC curves, but corroborated the unfavorable SOC points.
	The ROC curves for Other ICs were slightly degraded with respect to those in the five-class case, despite the SOC points remaining comparable. This could be due to the apparent difficulty in discriminating between Channel Noise ICs and Other ICs (sixth row of the ICLabel confusion matrix in Figure \ref{fig:cmat}). 
	
	Even though \icmsf{} had 10\% higher recall for Brain ICs than ICLabel in the five-class comparison, that gap nearly disappeared in the seven-class comparison. ICLabel's diminished recall of Brain ICs in the five-class case was likely a side effect of the approach used to merge classes. The summed probabilities of multiple, less probable classes can total to more than the probability of the maximal class in the unmerged comparison, possibly changing the IC classification of a single IC across the multiple comparisons. For example, while a label vector $\begin{bmatrix}0.45 & 0.4 & 0.15 \end{bmatrix}$ has maximal probability of belonging to the first class type, if the second and third classes are merged, the label vector becomes $\begin{bmatrix}0.45 & 0.55 \end{bmatrix}$ and the first class is no longer the most probable\footnote{
		This suggests an alternative means of performing the two-class and five-class comparisons: rather than first conflating the class probabilities through summation and then determining the maximal component, instead find the maximal IC category first and then combine the category labels. This method assures consistent discrete labels across varying numbers of IC categories. However, such a scheme prevents the use of measures dependent on predicted class probabilities such as cross entropy, ROC curves, and SOC points. It is for this reason that label conflation was performed as described in Section \ref{sec:eval}. Similar considerations are discussed further in Section \ref{sec:compositional_labels}.
	}.
	This only affected one and five ICs of the 130 total ICs for ICLabel\lite{} and ICLabel, respectively, when comparing the two-class and seven-class classifications.

	\subsection{IC Classification Speed}
	
	\begin{figure}
		\centering
		\includegraphics[scale=0.4]{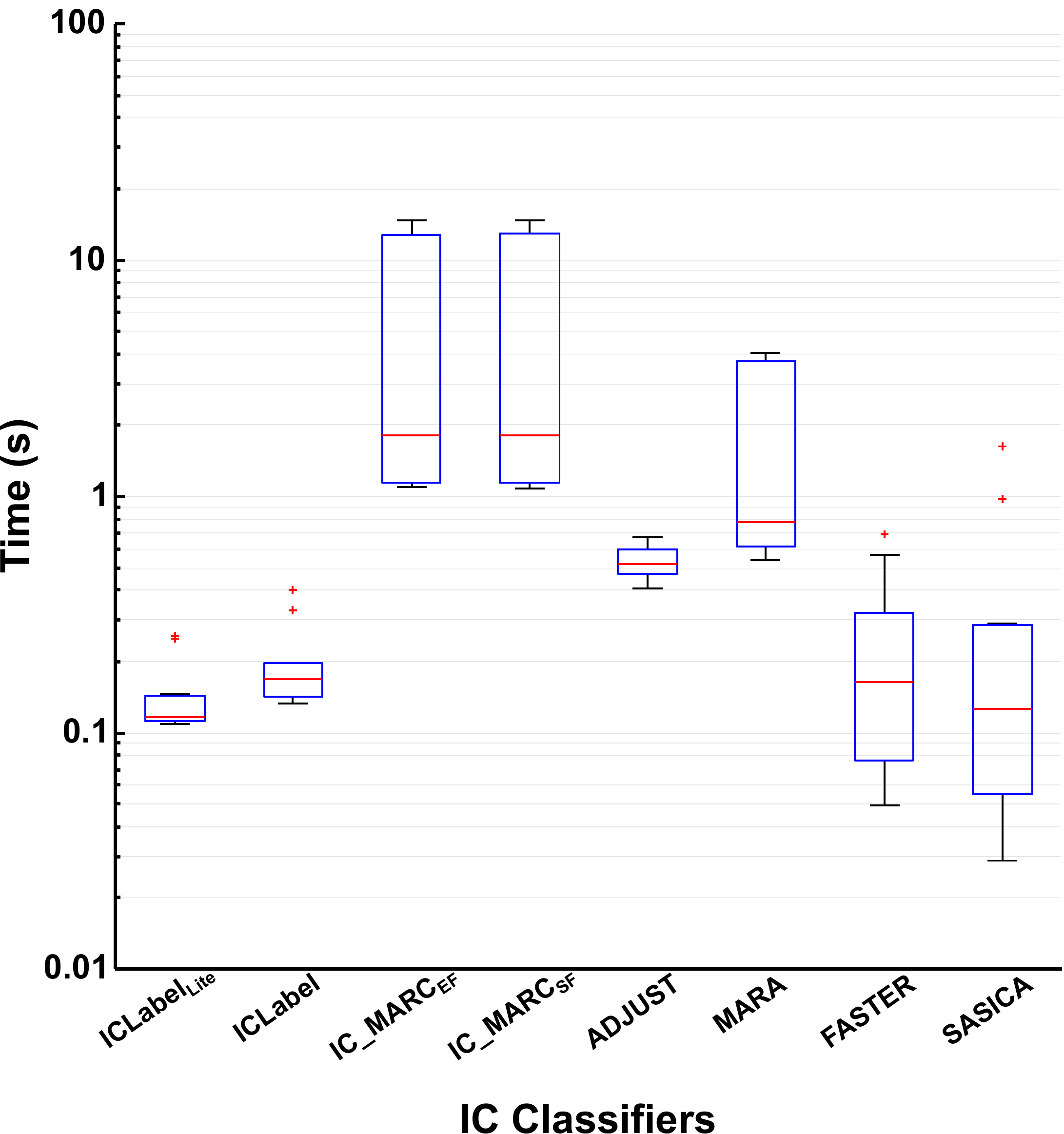}
		\caption{Time required to label a single IC, shown in logarithmic scale. Red lines indicate median time. Blue boxes denote the 25\textsuperscript{th} and 75\textsuperscript{th} percentiles, respectively. Whiskers show the most extreme values, excluding outliers which are denoted as small, red plus signs.}
		\label{fig:timings_log}
	\end{figure}
	
	Empirically-determined IC classification speeds can be found in Figure \ref{fig:timings_log}. Both IC\_MARC versions required similar run times: median 1.8 s per IC. ICLabel\lite{} and ICLabel required median run times of 120 ms and 170 ms respectively. These were (median) 15.5 and 13.0 times faster than IC\_MARC, respectively, and for single dataset averages up to a maximum of 88 and 64 times and a minimum of 9.8 and 6.7 times faster, respectively. Median IC classification speed for ICLabel\lite{} was 1.36 times faster than ICLabel, the difference required entirely due to the time taken to calculate the autocorrelation feature set. Details on the equipment used are provided at the end of \ref{app:metrics}.
	
	

	\section{Discussion}
	
	\subsection{Using Compositional IC Classifications}
	\label{sec:compositional_labels}

	\begin{table*}
		\centering
		\begin{tabular}{ccc|ccccccc}
			\toprule
			Classifier  & Dataset &       Metric       & Brain & Muscle & Eye  & Heart & L.N. & C.N. & Other \\ \midrule
			ICLabel    &  Train  & F\textsubscript{1} & 0.40  &  0.18  & 0.13 & 0.33  & 0.04 & 0.10 & 0.12  \\
			ICLabel    &  Train  &        Acc.        & 0.44  &  0.18  & 0.13 & 0.33  & 0.04 & 0.13 & 0.15  \\
			ICLabel    &  Test   & F\textsubscript{1} & 0.14  &  0.29  & 0.04 & 0.03  & 0.84 & 0.05 & 0.26  \\
			ICLabel    &  Test   &        Acc.        & 0.35  &  0.30  & 0.04 & 0.03  & 0.84 & 0.05 & 0.26  \\
			ICLabel\lite &  Train  & F\textsubscript{1} & 0.39  &  0.16  & 0.18 & 0.44  & 0.05 & 0.08 & 0.11  \\
			ICLabel\lite &  Train  &        Acc.        & 0.49  &  0.16  & 0.18 & 0.44  & 0.06 & 0.08 & 0.17  \\
			ICLabel\lite &  Test   & F\textsubscript{1} & 0.05  &  0.04  & 0.06 & 0.10  & 0.42 & 0.02 & 0.29  \\
			ICLabel\lite &  Test   &        Acc.        & 0.53  &  0.17  & 0.06 & 0.10  & 0.42 & 0.15 & 0.29  \\ \bottomrule
			\multicolumn{10}{c}{F\textsubscript{1}: F\textsubscript{1} Score; Acc.: Accuracy; L.N.: Line Noise; C.N.: Channel Noise} \\
			\bottomrule
			
		\end{tabular}
		\caption{Independent component (IC) category detection thresholds for multi-label classification under various conditions. Each set of thresholds was determined by selecting class-specific thresholds that maximized the specified metric on the specified datasets.} 
		\label{table:thresh}
	\end{table*}
	
	Compositional labels like those produced by ICLabel may be used in multiple ways.
	When a single, discrete label is required, as is typical for multi-class classification, compositional labels may be summarized by the category with maximal probability. When such an approach is taken, the value of the maximal probability can be interpreted as a measure of classifier confidence in the discrete classification.
	If the classification problem can be generalized to one of multi-label classification \citep{tsoumakas2007multi}, where each IC category is detected independent of other IC categories, each IC can be associated with zero or more different categorizations. In this case, class-specific thresholds can be applied to each IC category individually. 
	This method can leverage ROC curves to estimate optimal class-specific thresholds. The estimated optimal thresholds from the ICLabel training set and expert-labeled test set were determined by taking the point on each ROC curve with either maximal F\textsubscript{1} score or accuracy and are shown in Table \ref{table:thresh}. 
	Any element in a compositional IC label vector that matches or exceeds the corresponding threshold leads to a positive detection of the matching IC category. For example, using the thresholds determined from training set accuracy, if the ICLabel classifier produces an IC label vector $\begin{bmatrix} 0.71 & 0.04 & 0.03 & 0.01 & 0.01 & 0.02 & 0.18 \end{bmatrix}$, then the resulting detected labels would be $\begin{Bmatrix} \text{Brain}, & \text{Other} \end{Bmatrix}$ because $0.71 > 0.44$ and $0.18 > 0.15$.
	By comparison, when applying the multi-class classification approach of selecting the class with maximal associated label probability, the implicit threshold for detection could be any value between that of the maximum class probability and that of the next most probable class. Because of this variable threshold, which is effectively different for every example classified, classifier performance for discrete labels is harder to quantify using ROC curves, as each point on the curve is potentially relevant to classifier performance. In the multi-label case, ROC curves provide a direct performance estimate; when a single threshold is chosen, the classifier is reduced to a single point on the ROC curve and, therefore, has a single performance value in terms of TPR and FPR as defined in \ref{app:metrics}. 
	While multi-label classification is more flexible than multi-class classification, it allows for two possibly awkward outcomes: ICs with no IC category, and ICs with multiple IC categories. Depending on the use case, these outcomes may or may not be acceptable. 
	
	Compositional labels may also be used qualitatively to inform manual inspection. Compositional labels are more informative and easier to learn from than simple class labels \citep{hinton2015distilling}. They are also helpful for recognizing clearly mixed components by (1) showing which category is most likely applicable to an IC while also (2) indicating other IC types the component in question resembles. Compositional labels are also more informative in cases of classification error, by showing which other categories may be correct if the most probable one is not.
	While direct use of the compositional labels retains the most information provided by ICLabel, compositional labels may also be difficult to use in an automated fashion.
	

	\subsection{Timing}
	
	
	The speed of ICLabel feature extraction and inference theoretically allows the classifier to be used in online, near-real-time applications.
	Even though ICLabel\lite{} was typically 36\% faster than ICLabel, the average difference in calculation time per IC was only 50 ms. ICLabel is therefore sufficiently efficient for near-real-time use in most cases.
	A further consideration is that the times shown in Figure \ref{fig:timings_log} are based on features extracted from the entirety of each EEG recording. Those PSD and autocorrelation estimates are non-causal and thus impossible to actualize in the case of real-time applications. Instead, those features are best estimated using recursive updates that not only fix the issue of causality, but may also spread the computational cost of feature extraction across time. 
	By comparison, while IC\_MARC was claimed to be viable for real-time applications, the proposed paradigm in \citet{doi:10.1111/psyp.12290} consisted of offline ICA decompositions of three-minute data segments at three-minute intervals, providing for intermittently-updated solutions with delays of six minutes. Also, these times were provided with the explicit assumption of heavily parallelized computation. 
	
	An existing online application for ICLabel is in the Real-time EEG Source-mapping Toolbox (REST) \citep{pion2015real,pion2018online} which implements an automated pipeline for near-real-time EEG data preprocessing and ICA decomposition using online recursive ICA (ORICA) \citep{hsu2016real}. REST can apply an IC classifier in near-real-time to the ORICA-decomposed EEG data, either to select ICs of interest or reject specified IC categories. The retained ICs can be used to reconstruct a cleaned version of the EEG channel data in near-real-time.

	\subsection{Differences Between Cross-validated Training Data and Expert-labeled Test Set Results}
	
	ICLabel achieved higher scores on the cross-validated training data than on the expert-labeled test set. 
	This could have occurred for three possible reasons: (1) overfitting to the ICLabel training set, (2) differing labeling patterns between the crowdsourced training set and the expert-labeled test set, and (3) high variance in expert-labeled dataset performance measures owing to the relatively small size of that dataset (130 ICs) and relatively few designated expert labelers (6). Overfitting during training (1) is unlikely to have played a major role due to the combined use of early stopping and cross-validation \citep{amari1997cval} but factors (2) and (3) could both be contributing factors. To resolve either problem would require more labeled examples, especially examples labeled by experts \citep{della2012crowd}, a solution that is neither unexpected nor cheap. As more labels are submitted to the ICLabel website over time, these questions will become resolvable.

	\subsection{Cautions}
	As the primary purpose of an IC classifier is to enable automated component labeling, there is an implied trust in the results provided by that classifier. If the labels provided are incorrect, all further results derived from those labels are jeopardized. While the ICLabel classifier has been shown to generally provide high-quality IC labels, it is also important to be aware of its limitations, many of which are likely shared by other existing IC classifiers.
	
	The accuracy of the ICLabel classifier, like that of any classifier using a sufficiently powerful model, is primarily limited by the data used to learn the model parameters. While the ICLabel training set is large and contains examples of ICs from many types of experiments, amplifiers, electrode montages, and other important variables which affect EEG recordings, the dataset does not contain examples of all types of EEG data. 
	Infants, for example, are a population missing from the ICLabel dataset. As infant EEG can differ greatly from that of adults, spatially and temporally \citep{STROGANOVA1999997,MARSHALL20021199}, the results shown in Section \ref{sec:result-test} may not generalize to infant EEG. This issue was specifically raised by a user of the beta version of the ICLabel classifier who had anecdotal evidence of subpar performance when classifying Brain ICs in EEG datasets recorded from infants. While this is currently the only reported case of a possible structural failing of the classifier, more may exist relating to any other population of subjects or particular recording setting which is not sufficiently represented in the ICLabel dataset. Another likely source of datasets for which the ICLabel classifier could be unprepared is subjects with major brain pathology (brain tumor, open head injury, etc.). While recordings from subjects with epilepsy and children with attention deficit hyperactive disorder (ADHD) and autism are included in the ICLabel dataset, subjects with other conditions which might affect EEG may not be represented.
	
	Another concern is the quality of the electrode location data used to create the IC scalp topographies. Ideally EEG data should be accompanied by precise 3D electrode location data (now obtainable at low cost from 3D head images \citep{getchanlocs}), but the ICLabel dataset included some recordings that provided only template electrode location data, giving no simple means of controlling for localization error.
	
	
	%
	
	\subsection{An Evolving Classifier}
	The ICLabel project has the capacity to continue growing autonomously. Over time, as more suggested labels are submitted to the ICLabel website, automated scripts can perform the necessary actions of estimating ``true" labels using CL-LDA, training a new version of the ICLabel classifier, and publishing the new weights to the EEGLAB plug-in repository. To maintain consistency, there should then be three versions of the ICLabel classifier available in the EEGLAB plug-in: the automatically-updated classifier, the classifier validated here, and the early version of the classifier released to the public prior to publication of this article (ICLabel\iclbeta{}). While the individual segments of such a pipeline already exist, the overall automation is not yet in place and is therefore left as a future direction for the project.
	
	
	\section{Conclusion}
	The ICLabel classifier is a new EEG independent component (IC) classifier that was shown, in a systematic comparison with other publicly available EEG IC classifiers, to perform better or comparably to the current state of the art while requiring roughly one tenth the compute time. This classifier estimates IC classifications as compositional vectors across seven IC categories. The speed with which it classifies components allows for the possibility of detailed, near-real-time classification of online-decomposed EEG data. The architecture and training paradigm of the ICLabel classifier were selected through a cross-validated comparison between six candidate versions. A key component of the greater ICLabel project is the ICLabel website (\url{https://iclabel.ucsd.edu/tutorial}) which collects submitted classifications from EEG researchers around the world to label a growing subset of the ICLabel training set. The evolving ICLabel dataset of anonymized IC features is available at \url{https://github.com/lucapton/ICLabel-Dataset}. The ICLabel classifier is available for download through the EEGLAB extension manager and from \url{https://github.com/sccn/ICLabel}.

	\section{Acknowledgments}
	The expert-labeled test set was annotated with help from James Desjardins, Agatha Lenartowicz, Thea Radüntz, Lawrence Ward and Elizabeth Blundon, and Matthew Wisniewski. Their contributions are greatly appreciated. Thanks also to Francesco Marini for editorial comments.
	This work was supported in part by a gift from The Swartz Foundation (Old Field NY), by grants from the National Science Foundation [grant number GRFP DGE-1144086] and the National Institutes of Health [grant number 2R01-NS047293-14A1], and by a contract with Oculus VR, LLC. 
	Nvidia Corporation donated a Tesla K40 GPU through its GPU Grant Program which was used to efficiently train all artificial neural network models.
	
	\bibliographystyle{plainnat}
	\bibliography{Luca}
	
	\appendix

	\section{Evaluation Metrics} \label{app:metrics}
	
	\textbf{Balanced accuracy}, an average of within-class accuracies (within-class recall), 
	is defined as
	\[
	\frac{1}{C} \sum_{i=1}^{C} \frac{\rmtp_i}{\rmtp_i + \rmfn_i}
	\]
	where $C$ is the number of distinct classes and $\rmtp_i$ is the number of true positive detections, the number of correct classifications of examples into a specific class, for class $i$ and $\rmfn_i$ is the number of false negatives errors, the number of incorrect classifications of examples into any class other than the specific class, for class $i$. Although TP and FN are values that are typically calculated for binary classification, they can be easily adapted to the multi-class case by selecting one class as the ``positive" class and combining all other classes into the ``negative" class. In this way, $\rmtp_i$ is the number of correct classifications of examples into class $i$ and FN is the number of incorrect classifications of examples from class $i$ into any other class.
	
	\textbf{Cross entropy} is a measure that can be interpreted as the negative data log-likelihood if labels are assumed to be categorically distributed or alternatively as the portion of the Kullback–Leibler divergence that depends on predicted values. More pertinently, cross entropy was the primary metric optimized while training the ICLabel candidate classifiers, though it was modified for both the wCNN and GAN paradigms. 
	Cross entropy over an entire dataset is defined as 
	\[
	\sum_{n=1}^{N} \sum_{i=1}^{C} t_i^n \log p_i^n
	\]
	where N is the number of data-points and $t_i^n$ and $p_i^n$ are the $i$\textsuperscript{th} elements in the ``true" and predicted probabilistic label vectors, respectively, for the $n$\textsuperscript{th} IC.
	
	\textbf{The receiver operating characteristic (ROC) curve} shows the changing performance of a binary classifier as the threshold for detection of the positive class is varied from zero to one by plotting false positive rate (FPR) against true positive rate (TPR) on the horizontal and vertical axes, respectively. 
	TPR, also known as sensitivity or recall, is defined as $\rmtp / (\rmtp + \rmfn)$ which is the ratio of TP to total samples in the positive class.
	FPR is defined as $\rmfp / (\rmfp + \rmtn)$ where FP is the number of false positive errors, the number of incorrect classifications of examples into the positive class; TN is the number of true negative detections, that is, the number of correct classifications of examples into the negative class. FPR can also be defined as $1 - \text{specificity}$ where specificity is $\rmtn / (\rmfp + \rmtn)$. As was explained for balanced accuracy, one way ROC curves can be adapted to the multi-class case is by selecting a single class as the positive class and treating the combination of all other classes as the negative class.
	The ROC curve for the $i$\textsuperscript{th} class is a function of a threshold detection parameter $\theta \in [0,1]$ and is defined as the parametric function
	\[
	(\text{FPR}_i(\theta), \text{TPR}_i(\theta)) =
	\begin{cases}
	\text{TPR}_i(\theta) = \frac{\sum_{=1}^{N} \chi \left( p_i^n  \geq \theta \right) \chi \left( \argmax_k \; t_k^n = i \right)}{\sum_{n=1}^{N} \chi \left( \argmax_k \; t_k^n = i \right)} \\
	\text{FPR}_i(\theta) = \frac{\sum_{n=1}^{N} \chi \left( p_i^n  \geq  \theta \right) \chi \left( \argmax_k \; t_k^n \neq i \right)}{\sum_{n=1}^{N} \chi \left( \argmax_k \; t_k^n \neq i \right)} 
	\end{cases}
	\theta \in [0,1]\
	\]
	where $\chi(\cdot)$ is the indicator function defined as
	\[
	\chi \left( \texttt{condition} \right)  =
	\begin{cases}
	1 & \text{if \texttt{condition} is true} \\
	0 & \text{if \texttt{condition} is false}
	\end{cases}.
	\]
	When comparing threshold-dependent classifier performance on the ROC curve, ideal classifiers reside in the top left corner while a chance-level classifier resides along the diagonal connecting the bottom left and top right corners (see Figures \ref{fig:roc} and \ref{fig:roc_cv}). To aid in visual recognition of better curves, F\textsubscript{1} score isometrics are plotted that denote all point in the performance plane with equal F\textsubscript{1} score (higher value is better). The F\textsubscript{1} score is the harmonic average of recall and precision where precision is $\text{TP} / (\text{TP} + \text{FP})$ and the harmonic average of $x$ and $y$ is $1 / ((1/x) + (1/y)) = (xy) / (x+y)$. The F\textsubscript{1} score is convenient as it rewards reasonable compromises between precision and recall with higher values. For the experiments described earlier in this section, ROC curves are calculated for each IC category individually.
	
	\begin{figure}
		\centering
		\includegraphics[width=\textwidth]{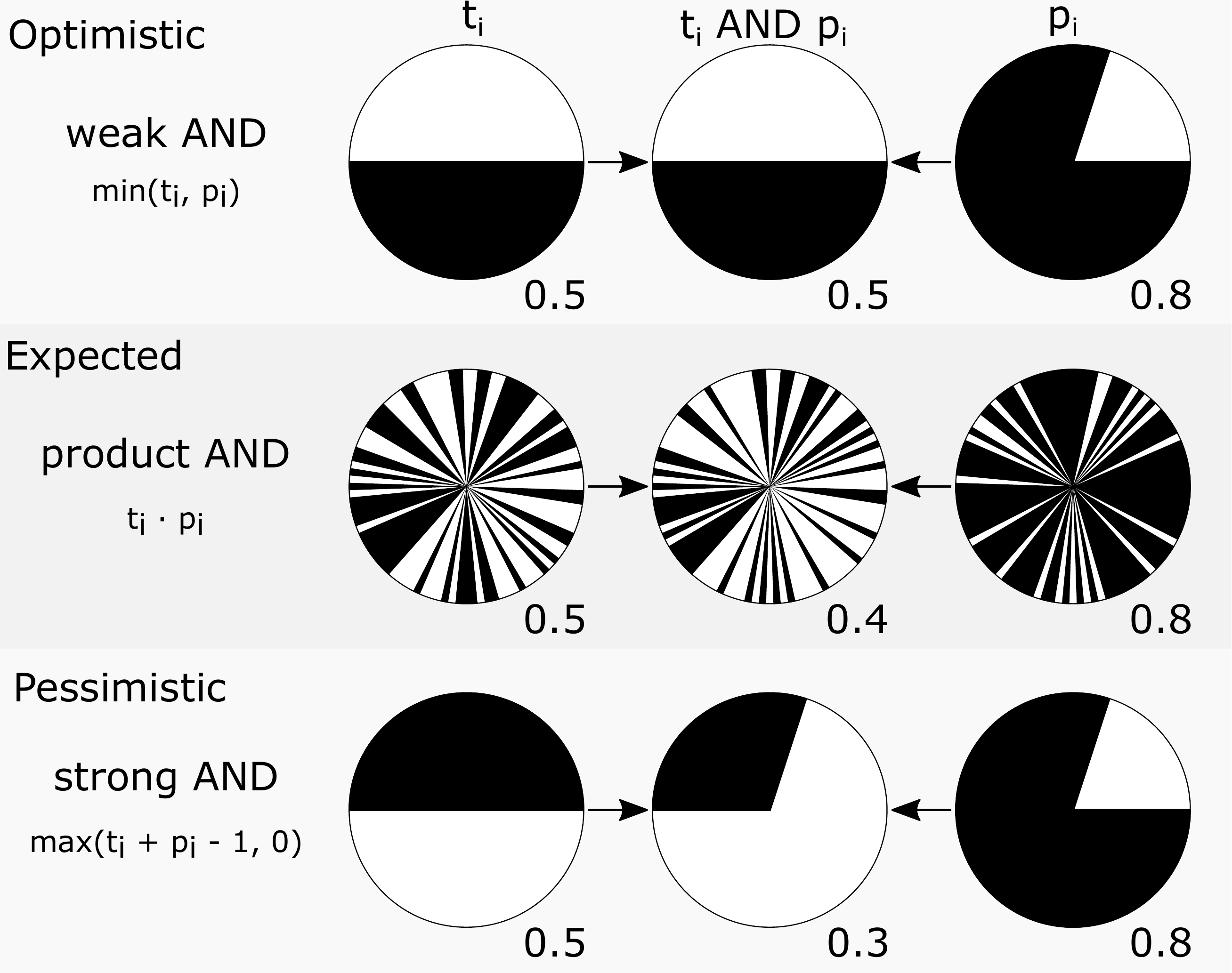}
		\caption{Visualization of three soft AND functions with which Boolean AND could be replaced for evaluating agreement between soft or compositional labels.
		The second and fourth columns from the left show how the reference and predicted class memberships (in black) might be distributed in a pie chart and the third row shows the resulting value of the Boolean AND of these soft-AND-related representative arrangements. 
	    Strong AND corresponds to the assumption of worst-case (lease) overlap of actual and predicted labels; expected AND corresponds to a uniform and independent distribution of actual and predicted labels; and weak AND corresponds to the best-case (most) overlap of actual and predicted labels.
	    The exact function related to each soft AND is given in the fourth row and the intuitive interpretation is given in the fifth row. This figure is modified after Figure 2 in \citet{BELEITES201312}.}
		\label{fig:soft_cm}
	\end{figure}
	
	\textbf{Confusion matrices} provide a matrix representation of the quantity and type of correct and incorrect classifications a classifier makes on a given dataset. As also explained in \ref{app:cllda}, each row is associated with a specific IC category determined through the crowd labeling effort, while each column is associated with a specific IC category as predicted by the classifier. Normally, the categories are in the same order for both the rows and the columns and therefore the diagonal elements are associated with true positive detections while the off-diagonal elements  are associated with errors. Normalized confusion matrices constrain the elements of each row to sum to 1 by dividing those elements by the total number of examples of each IC category. Mathematically, the elements of a normalized confusion matrix may be computed as
	\[
	\text{CM}_{ij} = \frac{\sum_{n=1}^{N} \chi \left( \argmax_k \; t_k^n = i \right) \chi \left( \argmax_k \; p_k^n = j \right)}{\sum_{n=1}^{N} \chi \left( \argmax_k \; t_k^n = i \right)}
	\]
	where $\text{CM}_{ij}$ is the element in the $i$\textsuperscript{th} row and the $j$\textsuperscript{th} column of the confusion matrix.
	
	\textbf{Soft confusion matrix} estimates account for the ambiguity of how soft labels and predictions might agree or differ \citep{BELEITES201312}. Rather than discretizing reference labels and predictions before counting how many match using the Boolean AND function, defined as
	\[
	\text{AND}(x, y) = 
	\begin{cases}
		1 & \text{if} \;\; x = y = 1 \\
		0 & \text{otherwise}
	\end{cases}
	\quad x, y \in \{0, 1\} \text{,}
	\]
	as for traditional confusion matrices, soft confusion matrices operate directly on continuous-valued soft label vectors and therefore require a different but comparable soft AND function for comparison. The aforementioned ambiguity in comparing soft labels arises from the various possible functions with which that comparison can be made. For example, assuming an IC contains activity from both the brain and line noise in equal proportions (i.e., 50\% ``Brain" and 50\% ``Line Noise", perhaps arising when the line noise activity was spatially non-stationary and therefore difficult to isolate through ICA decomposition), and that a classifier predicts that the IC is 20\% ``Brain" and 80\% ``Line Noise", three possible soft AND functions that can be used for comparison (strong AND, product AND, and weak AND) are detailed in Figure \ref{fig:soft_cm}. From an optimistic perspective, the ``Line Noise"-related agreement could be measured as the minimum of the two ``Line Noise"-related labels (weak AND) resulting in 50\% agreement as shown in the right-most column of Figure \ref{fig:soft_cm}. Alternatively the prediction of 80\% ``Line Noise" could have been wrongly based upon evidence originating from the brain-related aspects of the IC activity, therefore leaving only 30\% of the prediction being correctly derived from line-noise-related evidence. This pessimistic interpretation leads to the same result and interpretation as strong AND as shown in the second column from the left in Figure \ref{fig:soft_cm}. Weak AND and strong AND functions act as bounds on the possible ways that the labels and predictions conform and the actual agreement between label and prediction can be any value between those two, but assuming a uniformly distributed mapping of evidence to classifier prediction, the result would be 40\% agreement. This interpretation is associated with the product AND function and a visualization of such a uniform distribution of class-membership can be seen in the second column from the right in Figure \ref{fig:soft_cm}. This example is adapted from the cancer tissue example in Section 2.2 of \citet{BELEITES201312}, wherein this topic is more thoroughly explored.
	
	From these three continuous-valued replacements for the Boolean AND function, three different confusion matrices corresponding to pessimistic, expected, and optimistic estimates can be computed. These matrices can be combined to form pseudo-confidence intervals for elements of the soft confusion matrices and many of the statistics derived therefrom. Provided this fact, an equivalent to ROC curves, termed soft operating characteristic (SOC) points, may be computed by applying the TPR and FPR equations to the soft confusion matrices. As there is no discretization of the prediction in the soft case, the soft version of a class-specific ROC curve is only a single point per soft confusion matrix resulting in three total points in the performance plane per classifier and class. Following from the natural ordering of the strong, product, and weak AND functions, the three points making up each SOC are also ordered and are therefore connected by lines to show this relationship. Although soft-TPR and soft-FPR can be plotted on the same axes as classical ROC curves, the values along those the classical curves and the values derived from the soft confusion matrices are not directly comparable due to the conflicting assumptions guiding how each confusion matrix is calculated.
	
	The conclusion of \citet{BELEITES201312} lists four reason why a study might use soft confusion matrix statistics in place of the more commonly used statistics; these reasons are summarized here:
	\begin{enumerate}
		\item Label discretization, or ``hardening", leads to overestimating class separability.
		\item Estimating ambiguous labels may be a part of the goal for the predictor.
		\item Hardening explicitly disregards information present in the probabilistic labels.
		\item Hardening increases label variance when trying to learn smooth transitions between classes.
	\end{enumerate}
	Here, both ROC curves and SOC points are presented as the relevance of each measure depends on the intended application of a classifier.
	
	\textbf{IC Classification speed} was measured in terms of the time to extract features from and classify a single IC. The publicly available implementations of each classifiers was run, one dataset at a time, and the total calculation time for each dataset was divided by the number of ICs present in that dataset.
	This was repeated for all 10 datasets in the expert-labeled test set. Computations were performed in MATLAB on an AMD Opteron 6238 processor running at 2.6 GHz with no specified parallelization of calculations.
	
	\section{Generative Adversarial Networks} \label{app:gan}
	
	Generative adversarial networks (GAN) vie two competing artificial neural networks (ANN) against each other wherein one attempts to generate simulated data (generator network) and the other attempts to discern whether data is simulated or real data (discriminator network). Typically, GANs are trained in an a two-stage iterative fashion where in the first stage the generator network transforms random noise into simulated examples that the discriminator network classifies as either ``real" or ``fake". The generator network parameters are updated to make the discriminator more likely to label the generated examples as ``real". In the second stage, the discriminator labels another set of generated sample as well as actual collected samples. The discriminator network parameters are then updated to make the discriminator network more likely to label the generated samples as ``fake" and the actual samples as ``real". These two stages are repeated until predetermined convergence criteria are achieved.
	
	For SSGANs, instead of the discriminator network deciding between just real and simulated data, the ``real" category is subdivided into multiple classes such as ``Brain", ``Eye", and ``Other". The model used for the ICLabel classifier extended the SSGAN model to have multiple generator networks; one for each feature set used to describe ICs, that all shared the same random-noise input.
	As a final output, the SSGAN produced an eight-element compositional vector comprised of relative pseudo-probabilities for the seven IC categories described in Section \ref{sec:ictypes} and that of the IC being produced by the generator network. Regarding classification, the last element can easily be ignored by removing it and renormalizing the remaining seven-element vector to sum to one. 
	
	SSGANs have been shown to improve classification performance over CNNs when there are few labeled examples, provided there are more unlabeled examples available \citep{odena2016semi,salimans2016improved}. It has been theorized that the additional task of determining whether an example is real or generated helps the network to learn intermediate features helpful for classifying the examples into the categories of interest as well as discriminating actual from simulated ICs \citep{odena2016semi,salimans2016improved}. Others theorize that GANs help with classification when they generate low-probability examples that may be hard to find actual examples of in collected datasets. These low-probability examples help the network learn where the decision boundaries should be placed in the potentially large space between some classes \citep{dai2017good,lee2018training}, similar to the concept motivating maximum-margin classifiers like support vector machines. The training paradigms in \citet{dai2017good}, \citet{lee2018training}, and  \citet{srivastava2017veegan} were also attempted, but those results are omitted as they did not differ greatly from the modified SSGAN results shown in \ref{app:candidate_eval}.
	
	\section{ICLabel Candidate Classifier Selection} \label{app:candidate_eval} 
	
	\begin{figure}
		\centering
		\includegraphics[width=\textwidth]{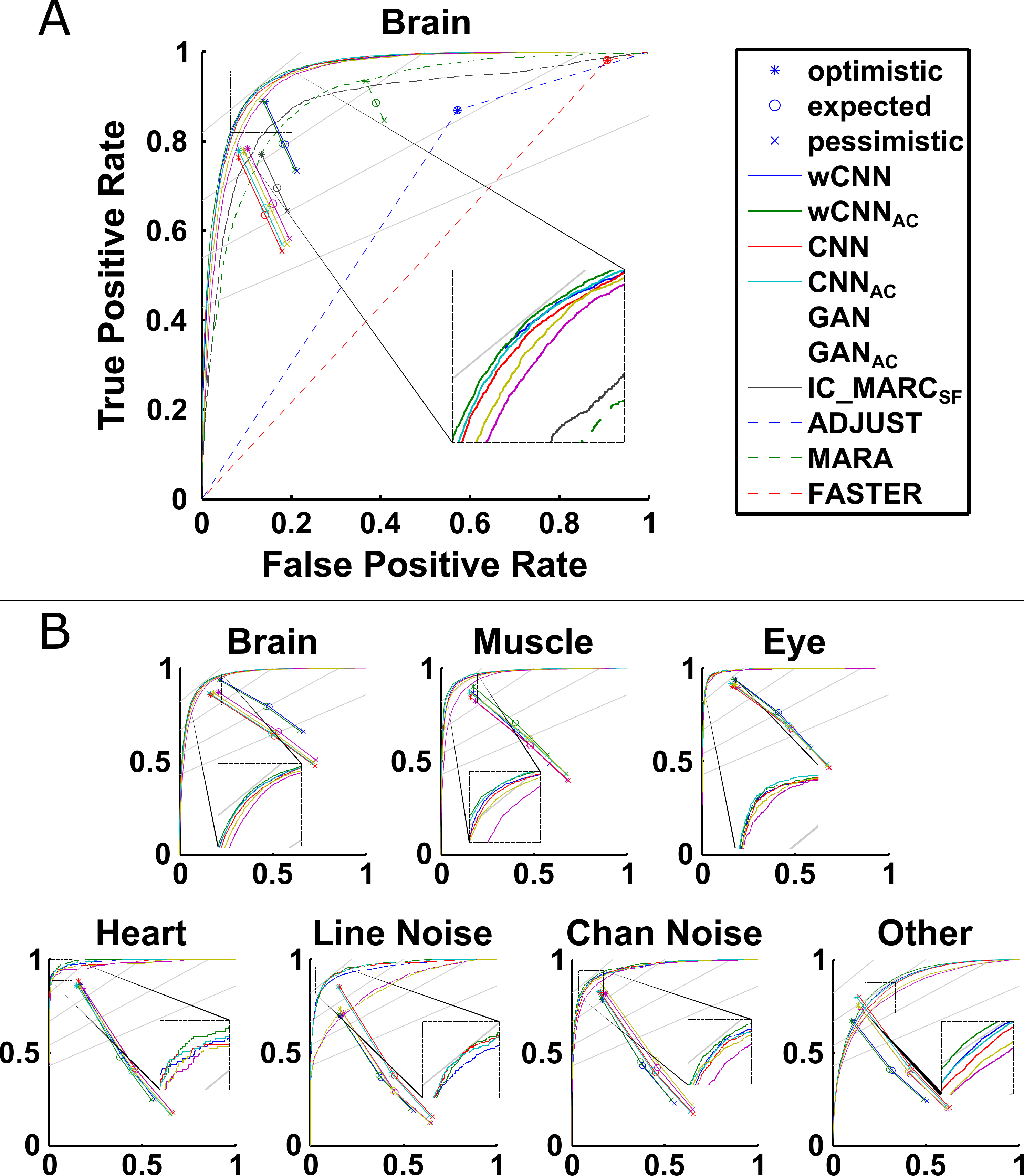}
		\caption{Color-coded ROC curves and soft operating characteristics (SOC) points calculated from soft confusion matrices to quantify IC classification performance on the cross-validated training data. The colors indicate the performances of the various candidate classifiers under consideration (see Sections \ref{sec:candidate} and \ref{sec:prior_methods} for the description of these classifiers). Part A of this figure contains the results merged into two classes, ``Brain" and ``Other", while part B contains the results across all seven ICLabel IC categories. The large dashed black squares show magnified views of the smaller dashed black squares. Gray lines indicate F\textsubscript{1} score isometrics of 0.9, 0.8, 0.7, and 0.6 from top to bottom. Refer to \ref{app:metrics} for definitions of F\textsubscript{1} score, ROC curves, and SOC points. The best performing candidate architecture was consistently shown to be wCNN\ac{}. The worst performing candidate architectures were those based on generative adversarial networks.}
		\label{fig:roc_cv}
	\end{figure}
	
	As described in Section \ref{sec:candidate}, six candidate IC classifiers were created in three-by-two factorial design to compare classification performance across three model architectures and training paradigms and two different collections of features provided to the candidate classifiers. These were measured using a ten-fold cross-validation scheme on the ICLabel training set.
	
	Regarding the first factor, model architecture and training paradigm, comparing ROC curves reveals that the GAN-based ICLabel candidates underperformed when compared to the other candidate models. This is visible across all seven classes in the ROC curves and most classes in the SOC points as presented in Figure \ref{fig:roc_cv}. 
	The exceptions for SOC points were ``Channel Noise" components, where the GAN methods scored highest on the soft measures, and Brain ICs and Eye ICs for which the GAN and unweighted CNN models performed similarly.
	While consistent, minor differences between wCNN and CNN models exist in the ROC curves, as shown for Other ICs and Chan Noise ICs, stronger differences are indicated by the SOC points where wCNN models notably outperformed CNN models. The wCNN models displayed better pessimistic and expected SOC performance over all classes as well as the best optimistic performance for Muscle ICs and Eye ICs. Despite exceptions in the case of Line Noise ICs and Other ICs, where the optimistic SOC points favored CNN models, the results generally favored wCNN models over CNN models. 
	
	For the second factor, feature sets provided to the candidate classifiers, the inclusion of autocorrelation as a feature set appeared to consistently improve performance across all classes. This was especially true for Muscle ICs and Other ICs, as evidenced by nearly uniform improvement measures by ROC curves and SOC points.
	
	With these three findings, the official ICLabel classifier was trained using the wCNN\ac{} paradigm and is referred to simply as ICLabel. This new model underwent comparison against published IC classification methods and, eventually, was publicly released as an EEGLAB plug-in. Because the autocorrelation feature set requires additional time to calculate, another model based on the wCNN paradigm was also compared with published IC classification methods for situations when faster feature extraction time is imperative. This new wCNN-based model is referred to as ICLabel\lite{}.
	
	\section{CL-LDA Details and Hyperparameters} \label{app:cllda}
	
	While reference labels (estimated ``true labels") are the desired output for the purposes of training the ICLabel classifier, CL-LDA also simultaneously calculates estimates of labelers’ skill, parameterized by a confusion matrix. For the ICLabel dataset, these confusion matrices take the form of seven-by-eight matrices where each row is associated with one of the seven IC categories mentioned in Section \ref{sec:ictypes} and each column is associated with one of the eight possible responses allowed on the ICLabel website: the seven IC categories and ``?". Each row of the confusion matrix can be interpreted as the estimated probabilities of the labeler providing each response conditioned on the IC in question being of that row's associated IC category. A perfect labeler would have ones in the entries for matching IC categories and responses, such as the intersection of the ``Brain"-response column and the Brain IC row, and zeros in the entries for mismatching IC categories and responses, such as the intersection of the Eye IC response column and the Brain IC row. These matrices start with prescribed values dependent on prior assumptions; but as labelers submit more labels, the labeler skill matrices become more dependent upon the submitted labels rather than those prior assumptions.
	
	CL-LDA efficiently estimates model parameters by maintaining counts of how each labeler labels examples from each IC category. In this way, priors on the labeler matrices can be interpreted as pseudo-counts that add their value to the actual, empirical counts tracked by CL-LDA. Compositional label estimates are formed by CL-LDA in much the same way using a weighted count of how labelers associate an IC with each IC category. Just as with the labeler priors, the class priors add pseudo-counts to the empirical counts for each IC. Refer to \citet{pion2017crowd} for more details. An implementation of CL-LDA can be found at \url{https://github.com/lucapton/crowd_labeling}. 
	
	Certain labelers were manually marked as ``known experts" when the ICLabel website database was created while the rest were treated as labelers of unknown skill. The experts were assigned a favorable and strong prior distribution for their confusion matrix parameters while the labelers of unknown skill were assigned a favorable and weak prior distribution of their confusion-matrix parameters. Strong and weak priors correspond to how many submitted labels are necessary to overcome that prior’s influence; strong requiring more and weak fewer. Explicit priors used in this work are provided below. To maintain an acceptable level of quality for labeler skill estimates, only labels from labelers who submitted ten or more labels were considered. If this requirement were not in place, there would be many votes included by users who submitted fewer labels and very little could be known regarding their abilities.
	
	The prior for expert confusion matrices was
	\[
	\begin{bmatrix}
	50.01 & 0.01  & 0.01  & 0.01  & 0.01  & 0.01  & 0.01  & 0.01 \\
	0.01  & 50.01 & 0.01  & 0.01  & 0.01  & 0.01  & 0.01  & 0.01 \\
	0.01  & 0.01  & 50.01 & 0.01  & 0.01  & 0.01  & 0.01  & 0.01 \\
	0.01  & 0.01  & 0.01  & 50.01 & 0.01  & 0.01  & 0.01  & 0.01 \\
	0.01  & 0.01  & 0.01  & 0.01  & 50.01 & 0.01  & 0.01  & 0.01 \\
	0.01  & 0.01  & 0.01  & 0.01  & 0.01  & 50.01 & 0.01  & 0.01 \\
	0.01  & 0.01  & 0.01  & 0.01  & 0.01  & 0.01  & 50.01 & 0.01
	\end{bmatrix}
	\]
	while the confusion matrix prior for labelers of unknown skill was
	\[
	\begin{bmatrix}
	1.25 & 0.25 & 0.25 & 0.25 & 0.25 & 0.25 & 0.25 & 0.25 \\
	0.25 & 1.25 & 0.25 & 0.25 & 0.25 & 0.25 & 0.25 & 0.25 \\
	0.25 & 0.25 & 1.25 & 0.25 & 0.25 & 0.25 & 0.25 & 0.25 \\
	0.25 & 0.25 & 0.25 & 1.25 & 0.25 & 0.25 & 0.25 & 0.25 \\
	0.25 & 0.25 & 0.25 & 0.25 & 1.25 & 0.25 & 0.25 & 0.25 \\
	0.25 & 0.25 & 0.25 & 0.25 & 0.25 & 1.25 & 0.25 & 0.25 \\
	0.25 & 0.25 & 0.25 & 0.25 & 0.25 & 0.25 & 1.25 & 0.25 \\
	\end{bmatrix}.
	\]
	Class priors were approximately
	\begin{equation*}
	\begin{bmatrix}
	0.002973 & 0.001766 & 0.00079 & 0.00015 & 0.000573 & 0.00073 & 0.003022 \\
	\end{bmatrix}.\\
	\end{equation*}
	
	The class priors were set as the empirically-determined class prior probabilities divided by 100 and are ordered following the same IC category ordering of the labeler confusion matrices. 
	The burn-in period for the CL-LDA Gibbs sampler was 200 epochs over the data and the labels were estimated over the next 800 epochs.
	
	To estimate labels for the expert-labeled test data, CL-LDA was applied to the collected expert labels on the test set using the same procedure as was used for the training set. The prior for expert confusion matrices was
	\[
	\begin{bmatrix}
	5    & 0.01 & 0.01 & 0.01 & 0.01 & 0.01 & 0.01 & 0.01 \\
	0.01 & 5    & 0.01 & 0.01 & 0.01 & 0.01 & 0.01 & 0.01 \\
	0.01 & 0.01 & 5    & 0.01 & 0.01 & 0.01 & 0.01 & 0.01 \\
	0.01 & 0.01 & 0.01 & 5    & 0.01 & 0.01 & 0.01 & 0.01 \\
	0.01 & 0.01 & 0.01 & 0.01 & 5    & 0.01 & 0.01 & 0.01 \\
	0.01 & 0.01 & 0.01 & 0.01 & 0.01 & 5    & 0.01 & 0.01 \\
	0.01 & 0.01 & 0.01 & 0.01 & 0.01 & 0.01 & 5    & 0.01
	\end{bmatrix}.
	\]
	and class priors were approximately
	\begin{equation*}\begin{bmatrix}
	0.002263 & 0.001537 & 0.001753 & 0.000155 & 0.00063 & 0.001839 & 0.001822 \\
	\end{bmatrix}.\\
	\end{equation*}
	
	\section{Artificial Neural Network Architecture Details} \label{app:arch}
	The ICLabel candidate and final classifiers were each composed of individual neural networks for each feature set, the outputs of which were concatenated and fed into another network to produce the final classifications.
	Specifically, the IC scalp topographies were fed into a two-dimensional CNN using dilated convolutions. One-dimensional CNNs were used for all other features (PSD and/or autocorrelation). 
	Scalp topography images were 32-pixels-by-32-pixels with one intensity channel. Both PSD and autocorrelation features sets were 100-element vectors. Scalp topographies and PSDs were scaled such that the maximum absolute value for each one was 0.99. Autocorrelation vectors were normalized such that the zero-lag value was 0.99 before removal.
	The discriminator and classifier scalp topography subnetworks were comprised of three convolutional layers 
	while the PSD and autocorrelation subnetworks had three one-dimensional convolutional layers. 
	The three generator subnetworks were comprised of four transposed convolutional layers each. As input, they took a shared 100-element vector of Gaussian noise with mean zero and a variance of one. This architecture was loosely based upon that of DCGAN \citep{radford2015unsupervised}.
	Details on the layers used in these architectures are shown in Table \ref{table:arch} where  ``Topo" is used as shorthand for scalp topography and ``AFC" for autocorrelation function. CNN and wCNN architectures only used layers in the ``Classifier" network, while GAN-based classifiers used all listed layers during training and only used ``Classifier" networks layers for inference.
	Classifier layer ``Final" used seven filters for both CNN and wCNN architectures while GAN-based classifiers used eight filters during training and seven during inference by removing the filter for detecting IC features created by the generator networks.
	GAN-based classifiers applied a binary mask to the output of the scalp topography generator network setting peripheral pixels to zero to match the interpolation format of actual scalp topographies.
	
	\begin{table}
		\centering
		\begin{tabular}{ccccccc}
			\toprule
			Network & Layer & Filters & Kernel & Stride & Padding & Activation \\
			\midrule
			Classifier & Topo-1 & 128 & 4$\times$4 & 2 & same & LReLU \\
			Classifier & Topo-2 & 256 & 4$\times$4 & 2 & same & LReLU \\
			Classifier & Topo-3 & 512 & 4$\times$4 & 2 & same & LReLU \\
			Classifier & PSD-1 & 128 & 3 & 2 & same & LReLU \\
			Classifier & PSD-2 & 256 & 3 & 2 & same & LReLU \\
			Classifier & PSD-3 & 1 & 3 & 2 & same & LReLU \\
			Classifier & ACF-1 & 128 & 3 & 2 & same & LReLU \\
			Classifier & ACF-2 & 256 & 3 & 2 & same & LReLU \\
			Classifier & ACF-3 & 1 & 3 & 2 & same & LReLU \\
			Classifier & Final & 7 or 8 & 4$\times$4 & 2 & valid & SoftMax \\
			
			Generator & Topo-1 & 2,000 & 4$\times$4 & 2 & valid & ReLU \\
			Generator & Topo-2 & 1,000 & 4$\times$4 & 2 & valid & ReLU \\
			Generator & Topo-3 & 500 & 4$\times$4 & 2 & valid & ReLU \\
			Generator & Topo-4 & 1 & 4$\times$4 & 2 & valid & tanh \\
			
			Generator & PSD-1 & 2,000 & 3 & 1 & valid & ReLU \\
			Generator & PSD-2 & 1,000 & 3 & 1 & valid & ReLU \\
			Generator & PSD-3 & 500 & 3 & 1 & valid & ReLU \\
			Generator & PSD-4 & 1 & 3 & 1 & valid & tanh \\
			
			Generator & ACF-1 & 2,000 & 3 & 1 & valid & ReLU \\
			Generator & ACF-2 & 1,000 & 3 & 1 & valid & ReLU \\
			Generator & ACF-3 & 500 & 3 & 1 & valid & ReLU \\
			Generator & ACF-4 & 1 & 3 & 1 & valid & tanh \\	
			\bottomrule
		\end{tabular}
		\caption{Layers used in ICLabel candidate classifier architectures.CNN and wCNN architectures only use layers in the ``Classifier" network, while GAN-based classifiers use all listed layers during training despite only using ``Classifier" networks layers during inference. Classifier layer ``Final" uses seven filters for both CNN and wCNN architectures while GAN-based classifiers use eight filters during training and seven during inference by removing the filter related to generated samples. ``Topo" is used as shorthand for ``scalp topography" and ``ACF" for ``autocorrelation function". ``ReLU" is short for ``rectified linear unit" \citep{relu2010}, ``LReLU" is short for ``leaky ReLU" \citep{maas2013rectifier} with a leakage parameter of 0.2., and ``tanh" is short for ``hyperbolic tangent".}
		\label{table:arch}
	\end{table}
	
	Training of the candidate and official models was accomplished using Adam \citep{kingma2014adam} with a learning rate of 0.0003, $\beta_1$ of 0.5, and $\beta_2$ of 0.999 to calculate parameter updates with a gradient cutoff of 20 and a batch size of 128 ICs. Labeled examples for each batch were selected with random class-balanced sampling to overcome class imbalances in the ICLabel training set. Holdout-based early stopping with a viewing window of 5,000 batches was used as a convergence condition to mitigate overfitting \citep{Prechelt2012}. All architectures used input noise \citep{DBLP:journals/corr/SonderbyCTSH16} to stabilize convergence. Batch normalization \citep{ioffe2015batch} was used only in the generator network from the GAN-based architecture. The GAN-based classifiers also used one-sided label smoothing \citep{salimans2016improved}.
	
	The ICLabel training set was augmented to exploit symmetries in scalp topographies through left--right reflections of the IC scalp topographies as well as negations of the IC scalp topographies. Negation of the scalp topography exploits the fact that if one negates both the ICA mixing matrix as well as the IC time-courses, the resulting channel data remain unchanged. As negating the time courses does not affect any of the other feature sets used, only the scalp topographies need be altered. Horizontal reflections of the scalp topographies exploits the (near) symmetry of human physiology. One notable exception to this symmetry is the heart being located only on the left side of the chest. However, Heart ICs were comparatively rare in the training set and left--right reflection of Heart IC scalp topographies did not create confusion with an other IC class scalp topography.
	This effectively resulted in a four-fold increase in the number of ICs in the dataset.
	
	All ICLabel candidate and official classifiers were built and trained in python using Tensorflow \citep{tensorflow2015-whitepaper}. They were also converted to MATLAB using matconvnet \citep{vedaldi15matconvnet} for distribution as an EEGLAB plug-in. Files involved in training the ICLabel classifier can be found at \url{https://github.com/lucapton/ICLabel-Train}.
	
\end{document}